\documentclass[aip,pof,reprint,unsortedaddress]{revtex4-1}
\usepackage{graphicx}
\usepackage{euscript,amscd}
\usepackage{latexsym}
\usepackage{amsmath}
\usepackage{amssymb}
\usepackage{mathrsfs}
\usepackage{array}
\usepackage{subfigure}

\usepackage{enumerate}

\usepackage{setspace}

\def\Rt{Re_{\tau}}

\def\pa{\partial}
\def\ov{\overline}

\def\Ah{\hat{A}}
\def\Bh{\hat{B}}
\def\Ch{\hat{C}}
\def\Dh{\hat{D}}
\def\Ph{\hat{P}}
\def\Ab{\bar{A}}
\def\Bb{\bar{B}}
\def\Cb{\bar{C}}
\def\Db{\bar{D}}
\def\Pb{\bar{P}}
\def\At{\tilde{A}}
\def\Bt{\tilde{B}}
\def\Ct{\tilde{C}}
\def\Dt{\tilde{D}}

\def\Gt{\tilde{G}}
\def\Qt{\tilde{Q}}
\def\Kt{\tilde{K}}

\newcommand\x[1]{{x_{#1}}} 
\newcommand\scale[1]{{\lambda_{#1}}}

\def\Um{\overline{U}} 
\def\Ul{U_0} 

\newcommand\um[1]{{\tilde{u}_{#1}}}
\newcommand\pmt{\tilde{p}}
\newcommand\vortm[1]{{\tilde{\eta}_{#1}}}

\newcommand\ul[1]{{u_{#1}}}
\newcommand\pl{p}

\def\rank{\text{rank}}

\def\BibTeX{{\rm B\kern-.05em{\sc i\kern-.025em b}\kern-.08em
    T\kern-.1667em\lower.7ex\hbox{E}\kern-.125emX}}

\def\Hinf{$\mathcal {H}_\infty~$}
\def\hinf{$\mathcal {H}_\infty~$}
\def\htwo{${\mathcal {H}}_2~$}
\def\Htwo{${\mathcal {H}}_2~$}

\def\ie{\emph{i.e.~}}

\def\s={\stackrel{s}{=}}

\newcommand{\curly}[1]{{\mathcal{#1}}}

\newcommand{\bspace}[1]{{\mathbb{#1}}}
\newcommand{\real}[1]{{\text{Re}}({#1})}

\newcommand{\pderiv}[2]{\frac{\partial #1}{\partial #2}}
\newcommand{\pderivsq}[2]{\frac{\partial^{2} #1}{\partial #2^{2}}}

\newcommand{\norm}[2]{\lVert#1\rVert_{#2}}
\newcommand{\inprod}[2]{\left<{#1},{#2}\right>}

\newcommand{\mat}[2]{\left[\begin{array}{#1} #2\end{array}\right]}

\begin{document}


\title{Relaminarisation of $Re_\tau=100$ channel flow with globally stabilising linear feedback control} 



\author{AS Sharma}
\email[]{a.s.sharma @ sheffield.ac.uk}
\homepage[]{http://www.sheffield.ac.uk/acse/staff/sharma}
\affiliation{Department of Automatic Control and Systems Engineering, University of Sheffield}
\author{JF Morrison}
\email[]{j.morrison@imperial.ac.uk}
\affiliation{Department of Aeronautics, Imperial College}
\author{BJ McKeon}
\affiliation{Graduate Aeronautical Laboratories, California Institute of Technology}
\author{DJN Limebeer}
\affiliation{Department of Engineering Science, University of Oxford}
\author{WH Koberg}
\affiliation{Department of Aeronautics, Imperial College}
\author{SJ Sherwin}
\affiliation{Department of Aeronautics, Imperial College}


\date{\today}

\begin{abstract}
The problems of nonlinearity and high dimension have so far prevented a complete solution of the control of turbulent flow.  Addressing the problem of nonlinearity, we propose a flow control strategy which ensures that the energy of any perturbation to the target profile decays monotonically. The controller's estimate of the flow state is similarly guaranteed to converge to the true value. We present a one-time off-line synthesis procedure, which generalises to accommodate more restrictive actuation and sensing arrangements, with conditions for existence for the controller given in this case. The control is tested in turbulent channel flow ($Re_\tau=100$) using full-domain sensing and actuation on the wall-normal velocity. Concentrated at the point of maximum inflection in the mean profile, the control directly counters the supply of turbulence energy arising from the interaction of the wall-normal perturbations with the flow shear. It is found that the control is only required for the larger-scale motions, specifically those above the scale of the mean streak spacing. Minimal control effort is required once laminar flow is achieved. The response of the near-wall flow is examined in detail, with particular emphasis on the pressure and wall-normal velocity fields, in the context of Landahl's theory of sheared turbulence.
\end{abstract}


\maketitle 

\section{Introduction}

The return of turbulent wall flow to the laminar state is a problem with diverse and important applications, such as those in the aeronautics, shipping and oil industries. A comprehensive solution of the flow control problem still faces serious challenges. The main obstacles to a complete flow control theory are that the governing equations are nonlinear and of infinite dimension. This paper aims to address the problem of the nonlinearity, and in doing so, improve the understanding of the physical processes. As a consequence, some possible approaches to the problem of high dimensionality present themselves.

In addition to the nonlinearity, flow control strategies must deal with model uncertainty and the exogenous disturbances such as vibration and free-stream disturbances that arise in realistic applications.
This would suggest the use of closed-loop control strategies, which involve the feeding back of measurements of the system output into current and future control decisions, over open-loop strategies, where the actual system output is not compared to the desired output. Of the two classes, closed-loop control, offers superior robustness characteristics in the face of modelling error, state uncertainty and exogenous disturbance.

\Hinf control theory has been notably successful in providing good control performance for systems with the large class of bounded uncertainties, nonlinearities and exogenous disturbances \cite{Green, Doyle89}. Importantly for us, the \Hinf theory can be generalised to cope with the nonlinearity we face with the Navier-Stokes equations.

Modern control methods typically make assumptions about the kind of model error or disturbance that is present in the system. \Htwo or optimal control assumes Gaussian state and measurement disturbances and \Hinf control typically gives stability guarantees for model errors up to a certain bound. 
The Navier-Stokes nonlinearity is neither stochastic nor bounded.  However in a closed or periodic domain, it is well known to be conservative with respect to the perturbation energy. In the following, we exploit this fact using the passivity theorem. The resulting control gives global stability guarantees (and consequently relaminarisation) for the discretised, controlled Navier-Stokes flow, where actuation and measurement requirements are met. This approach has been proposed\cite{Sharma05APS} and outlined\cite{Sharma06AIAA} in previous work. Results of our approach applied to simulations of turbulent flows are presented here for the first time, with application to turbulent channel flow.

From a control perspective, this work advances on previous work in at least three important respects. First, the current approach is demonstrated for a turbulent, three-dimensional flow. Second, it offers a constructive synthesis procedure. Third, it provides limits on the turbulent energy production where global stabilisation is not possible due to insufficient actuation or sensing.

With reference to known existing theory, the work also describes the physics behind the controller action and explains why a linear control strategy is always sufficient to attenuate turbulence.

\subsection{Modern flow control}
A comprehensive review of modern feedback flow control is available in the paper by Kim and Bewley\cite{Kim07}. Further physical insight is given in the recent paper by Kim\cite{Kim11}.
\Htwo (``optimal'') and \Hinf (``robust'') designs have been applied to the linearised transition delay problem for particular wavenumber pairs by Bewley and Liu\cite{Bewley98}. The optimal control approach was tried first by Joslin\cite{Joslin97}. Joshi {\it et al.}\cite{Joshi97} have proposed a simple controller design using classical methods which was used with some success in the stabilisation of infinitesimal and finite-amplitude disturbances. H\"ogberg {\it et al.}\cite{Hoegberg03} have demonstrated that linear feedback control can be used to increase the threshold perturbation amplitudes for transition to occur. In another work, H\"ogberg {\it et al.}\cite{Hoegberg03-1} presented a gain-scheduling approach which  relaminarised turbulent flows in all instances tested, however no proof of global stability was offered. Further results on linear flow control methods are available in the book by \r{A}mo and Krsti\'{c}\cite{Aamo}. The paper by Fukagata {\it et al.}\cite{Fukagata05} investigates interior forcing targeting the Reynolds stress terms directly.

These \Htwo and \Hinf linear flow control strategies have been designed to delay transition, by preventing the flow from leaving the regime of small perturbations to the desired laminar flow. Given the likelihood in practical situations of large excursions due to transients, exogenous disturbances, or model error, the assumption of small perturbations seems inappropriate.

In contrast, the remarkable feature of our approach is that a \emph{linear} synthesis problem provides a \emph{global}, \emph{nonlinear} stability result. An implicit consequence of the stability result is the convergence of the estimation problem. For the first time, the controller's internal representation of the flow is guaranteed to converge to the true value. In contrast to the \Htwo setting, the control and estimation problems are inexorably coupled and the ``separation principle''\cite{Green} does not apply.

Perhaps most pertinent to the current approach is the work described by Balogh, Liu and Krsti\'{c}\cite{Balogh01}. Their approach uses a Lyapunov stability argument (comparable to the passivity argument used here) to prove the existence of a globally stabilising linear control, using relatively realistic tangential boundary actuation and shear stress measurements. A proof of global asymptotic stability is offered for the two-dimensional case and it is claimed that the method will work in three dimensional flows. Their proof is applicable only at ``sufficiently low'' Reynolds numbers.

\subsection{An overview of the passivity approach}
Conceptually, the method of control that we propose is simple, although the details of the synthesis are more complicated.
The idea is to eliminate perturbations to a laminar flow, which we take as the operating point of the system. We do not model fluctuations about the mean profile. As such, the difference between the turbulent mean and the laminar profile would be included in the perturbations that we wish to decay.

The flow at any particular wavenumber experiences a coupling from other wavenumbers, via the nonlinear convective term, that we consider as a forcing. This nonlinear term does not produce or destroy perturbation energy. Consequently, if by means of feedback control, the system at every wavenumber can be made to dissipate energy, then the system as a whole will still dissipate energy, even with the nonlinearity. This objective involves only the linear system at each wavenumber and can therefore be solved using linear synthesis methods.

There is a direct analogy from circuit theory. The energy in any circuit made up of passive components like capacitors, resistors, inductors, will always decay, in the absence of a non-passive component such as a battery.

In the case of turbulent flow, relaminarisation could be achieved simply by introducing a large dissipative term, analogous to a resistor. Such an approach of adding more viscosity cannot generalise to other types of actuation, and so is not followed.

Henningson \& Reddy\cite{Henningson94} showed that non-normality of the system matrix governing perturbations to the laminar flow solution is a necessary condition for subcritical turbulence; hence, imposing linear stability \emph{and} destroying the non-normality by means of feedback control provides a sufficient condition for laminar flow.

That this condition is linear greatly simplifies the synthesis procedure, allowing use of the superposition principle, so that we can find the controller at each Fourier mode separately, without sacrificing the nonlinear stability result. The controller may be found once, off-line, since it applies at all states at a given Reynolds number. Since the controller is linear, the most complicated mathematical operation it must perform on-line is a small number of matrix multiplications at each time step. This stands in contrast to nonlinear adjoint-based approaches and has obvious beneficial implications for eventual practical implementation. With the further application of model- or controller-reduction methods, the computations at this step could be reduced further.

Although many results for finite-dimensional systems in the control literature have analogues in the infinite-dimensional setting, the finite-dimensional theory is usually much simpler and is computationally tractable. This is particularly true for the case of the Navier-Stokes equations, where proofs of even rudimentary properties remain elusive\cite{Doering09}. When the system equations are discretised, as is typically done for practical control problems, the finite-dimensional control theory also provides a controller synthesis procedure. The procedure we use is described in the literature\cite{Safonov87} and involves a number of transformations of the linearised equations, then the solution of two Algebraic Riccati Equations (AREs). Whether solutions to these AREs exist depends on the actuation and measurement capabilities available to the control algorithm. Because the conditions for existence of AREs are well understood and easily checked, this problem formulation can inform the system designer about the suitability of proposed sensor and actuation arrangements at the design stage.

The passivity requirement is a conservative control strategy and may not be achievable with restricted actuation or sensing capability. To handle this case, the synthesis methodology presented iteratively approaches the ideal case and offers bounds on the worst-case perturbation energy production permitted by the controlled flow.
Because we may relax the control objective of passivity, we envisage that the current framework may be applied to more restrictive types of body forcing.

For this study, full-field volume forcing and measurement of wall-normal velocity only is applied to periodic, turbulent channel flow at $Re_\tau=100$. This simplified forcing and measurement arrangement is chosen to avoid confusion between the relative importance of various physical effects and the choice of any particular type of actuation.

\subsection{Linear processes and structure in wall turbulence}
The importance of linear mechanisms in turbulence has been understood since Batchelor and Proudman\cite{BP54} put forward their theory of rapid distortion (for a review see Hunt and Carruthers\cite{HuntC90}). Phillips\cite{phillips69} has suggested that, in shear flows, the Reynolds stresses arise from the direct interaction between the turbulence and the mean shear rather than a result of indirect, nonlinear interactions.  This implies that linear control schemes taking advantage of these mechanisms may be successful in attenuating turbulence.

Lee, Kim and Moin\cite{LeeKM90} have shown how many of the important dynamical processes are captured by rapid distortion theory (RDT) and comparison with direct numerical simulation (DNS) of sheared homogeneous turbulence shows that, despite its linear approximations, RDT retains ``the essential mechanism for the development of turbulence structures in the presence of high shear rate typical of the near-wall region in a turbulent shear flow''.  The qualitatively similar theories of Ellingsen and Palm\cite{EllingP:pof75} and Landahl\cite{landahl80:jfm,landahl90:jfm} offer an explanation for the importance of three-dimensional disturbances in inviscid, parallel shear flow: the latter offers an explanation for the formation and lift-up of near-wall streaks by an algebraic instability and the present work is interpreted in the context of Landahl's ideas.  The importance of the theory comes from the fact that it is both linear and is based on an analysis of the inviscid Rayleigh equation, suggesting that it is relevant to dynamical processes at any Reynolds number.

Non-normality appears as a recurrent theme in the literature on transition (Schmid and Henningson\cite{schmid:henningson}, Schmid\cite{Schmid07}). Butler and Farrell\cite{Butler92} investigated initial conditions which are capable of the greatest energy growth (which they call ``global optimal perturbations'') at a given Reynolds number.  They found that, in plane Poiseuille flow, the global optimal perturbation consists of a pair of counter-rotating streamwise vortices, though other modes should not be underestimated: in particular, oblique modes grow less, but faster. It remains an open question how important the initial condition problem \emph{per se} is in flows that are already turbulent.

Farrell and Ioannou\cite{Farrell93, Farrell96} have suggested that the linearised Navier-Stokes equations in plane channel flow under stochastic forcing can exhibit behaviour reminiscent of the streamwise vortices and streaks characteristic of turbulent flow. Kim and Lim\cite{Kim00} demonstrated in simulations of turbulent channel flow that the turbulence decays without the term coupling the wall-normal vorticity and the wall-normal velocity in the linearised Navier-Stokes equations. Henningson and Reddy\cite{Henningson94} have shown that non-normality is a necessary condition for sub-critical transition, i.e.~the linearised Navier-Stokes equations must have either exponentially growing modes or transiently growing solutions for transition to occur.

The prevalence of streaks and quasi-streamwise vortices in near-wall turbulence has been known for some time (Kline {\it et al.} \cite{kline:etal67}, Kim {\it et al.}\cite{kim:etal71}, Robinson \cite{robs91}), though which feature causes the other is still a subject of discussion (see for example Chernyshenko\cite{Chernyshenko05}). More recently, Kim \& Adrian\cite{kim:adrian}, Ganapathisubramani {\it et al.}\cite{ganapath06}, Hutchins \& Marusic\cite{hutchins07} and Guala {\it et al.}\cite{guala06} have shown the importance of very large scale motions (VLSMs or ``superstructures'')  which carry approximately half the Reynolds stress. Recently, they have been shown to appear on very rough surfaces also (Birch and Morrison\cite{birchm:jfm11}).  The description of a streak lift-up (burst) as an ``instability'' initiated by the large-scale disturbances from the outer layer appears in the seminal papers of Kline {\it et al.}\cite{kline:etal67} and Kim {\it et al.}\cite{kim:etal71}.  Morrison\cite{morrison:philtrans} describes more recent ideas concerning inner-outer interaction and its relationship to ``inactive motion'' \cite{a2t:jfm61,pb:jfm67,a2t:book76}.

Hall \& Sherwin\cite{Hall10} take an alternative approach of considering inviscid waves in the wavy critical layer (where the wave speed is close to the convective velocity) of a streaky base flow. Using an earlier theory due to Hall \& Smith \cite{Hall91} on vortex/wave interaction, they describe the nonlinear interaction of a self-sustaining process in which the nonlinear terms of finite-amplitude waves drive the streamwise vortices through a jump in the stresses at the critical layer, which in turn drive the streaks present in the base flow.
The unstable equilibrium solution they determine describes the whole cycle of the self-sustaining process and can be loosely considered as a nonlinear eigenvalue problem.
As such, it may be kept in mind where we use the description of a linear system driven by a nonlinear feedback forcing.

Following Reynolds and Hussain\cite{ReynoldsHussain3:72}, del {\'A}lamo and Jim{\'e}nez\cite{delAlamo06} have undertaken a temporal stability analysis of the Orr-Sommerfeld and Squire (OSS) equations in turbulent channel flow ($Re_{\tau}=2000$) using a variable eddy viscosity: they show that maximum amplification of disturbances occurs at two spanwise wavelengths, one corresponding to the widely accepted streak spacing, $\lambda_\x{3}^+=100$, ( the $~`+'$ superscript denotes a variable non-dimensionalised by the viscous length scale, $\nu/u_{\tau}$ where $\nu$ is the kinematic viscosity and $u_\tau$, the friction velocity, $u_\tau = \sqrt{\tau_w / \rho}$, where $\tau_w$ is the wall shear stress and $\rho$ is the density), the other occurring at $\lambda_\x{3}=3h$, where $h$ is the channel half-height.  While the former is clear evidence of near wall streaks (with streamwise wavelength of $\lambda_\x{1}\approx 1000$), the latter indicates the presence of VLSMs.  They also note that the fluctuations in streamwise velocity contain nearly all the kinetic energy and last longer than those in the wall-normal velocity.  In a similar vein, Hwang and Cossu\cite{cossu10:jfm} have shown that, in a turbulent channel flow for a sufficiently large Reynolds number, two distinct peaks of optimal growth appear, one scaling with viscous scales, the other with outer scales.

McKeon and Sharma\cite{McKeon10} have explored a simple, essentially linear, forcing-response type description of the dominant processes in high-Reynolds-number turbulent pipe flow. The model reproduces inner scaling of the small scales close to the wall and outer scaling in the flow interior and displays features representative of VLSMs including their modulation of the smaller scale features. In contrast to Landahl's theory, the work has addressed scaling with Reynolds number. However, both theories stress the importance of linear mechanisms. One important feature of the theory is that a high response to forcing is observed around the critical layer, and in regions of high shear. At higher Reynolds numbers, the theory predicts that the effect on the near-wall turbulence of VLSM-type structures become more important, whereas at lower Reynolds numbers the lift-up mechanism is more important. This prediction is supported by the control simulations of Touber \& Leschziner\cite{Touber11}.

\subsection{Passivity}

Passivity is an energy concept; its origin lies in circuit theory\cite{Wyatt81}. A component is called passive if only a finite amount of energy may be extracted from it.
To take a physical example, if $f(t)$ is a forcing function or field at time $t$ on a system $Z$ and $v(t)$ is its velocity, such that $v=Zf$, then the power consumed by the system from time $t=0$ to time $T$ is $\left< f, v \right>_{[0,T]} = \int_0^T f(t) v(t) ~dt$.
If we assume the initial conditions are zero, and this integral is positive for all $T$, then the system is passive. Essentially, it is a statement that the instantaneous power consumption is always positive.
If $Z$ is linear, the requirement for $\left<f,v\right>_{[0,T]}>0$ is then equivalent to the requirement that $Z$ is positive real, $Z(j\omega) + Z^*(j\omega)>0$. To show this, the integral $\left<f,v\right>_{[0,T]} = \left<f,Zf\right>_{[0,T]}$ is considered in the frequency domain.

The passivity theorem simply states that the feedback interconnection of two passive elements is itself passive. Intuitively put, if two elements which cannot produce energy are connected in a feedback arrangement, then the feedback arrangement as a whole also cannot produce energy. We will show this for our particular case, and for general proofs and more information the reader is directed to standard control texts\cite{Green, vdSchaft}.

\subsection{Landahl's theory, scales and waves}
It is instructive to preface the current analysis with a review of the basic ideas of Landahl's theory\cite{landahl67:jfm,landahl75:siam,landahl77:pof,landahl90:jfm,landahl93:eurjmech}, as inspired by the early wave theories of the viscous sublayer (Sternberg\cite{sternberg62} and Morrison, Bullock and Kronauer\cite{mbk71}). Here we define a wave as a motion that exhibits a convection velocity that is constant over a region in wall-normal distance. An accepted definition of wave motion is one in which energy is transported but without bulk motion: hence the wave motion refers only to the fluctuating pressure and velocity field.  Note that this is a stronger requirement (and a more physical definition) than a superposition of Fourier modes (Phillips\cite{phillips69}).  Therefore defining the viscous sublayer as a wave guide in which the least-damped waves exhibit significant correlation over large distances\cite{landahl67:jfm}, while useful, should not be taken too far.

The current approach is reminiscent of Landahl's ideas, because both theories have the nonlinear terms as a right-hand side forcing to the linear problem as a common point of departure. However Landahl's theory considers perturbations to the mean turbulent profile, whereas we will consider perturbations about the laminar one. This difference will be examined more fully below. Specifically, Landahl's theory\cite{landahl75:siam,landahl77:pof,landahl90:jfm} considers the Orr-Sommerfeld and Squire equations, which respectively, may be written as
\begin{equation}
\frac{\overline{D}\nabla^2 \um{2}}{D t}-\Um^{''}\frac{\partial \um{2}}{\partial \x{1}}-\frac{\nabla^4 \um{2}}{Re}=q
\label{orrsom}
\end{equation}
\begin{equation}
\frac{\overline{D}\vortm{2}}{D t}+\Um^{'}\frac{\partial \um{2}}{\partial \x{3}}-\frac{\nabla^2 \vortm{2}}{Re}=r,
\label{squire}
\end{equation}
where
\begin{equation}
\frac{\overline{D}}{D t}=\left(\frac{\partial}{\partial t}+\Um\frac{\partial}{\partial \x{1}}\right),
\end{equation}
$\bar{U}$ is the mean flow profile $\ov{U}(y)\delta_{i1}$ subject to a three-dimensional disturbance with velocity $\um{i}(\x{j}, t)$ and pressure, $\pmt(\x{i},t)$. Here the streamwise direction is $\x{1}$, the wall-normal direction is $\x{2}$ and the spanwise directions is $\x{3}$, giving the total velocity field as $U_i(x_j,t)=\Um(y)\delta_{i1}+\um{i}(\x{j},t)$. The wall-normal vorticity is $\vortm{}=\nabla \times \um{}$. The forcing terms $q$ and $r$ are quadratic terms involving Reynolds stresses. Landahl proposed that $q$ and $r$ are significant only in localised regions in space and time, thus giving rise to ``compact" source terms in (\ref{orrsom}) and (\ref{squire}) (see for example Landahl\cite{landahl75:siam}). The picture of sublayer motion is therefore one in which regions of ``intense small-scale turbulence of an intermittent nature" are interspersed by periods of ``laminar-like but unsteady motion of larger scale".

Landahl\cite{landahl93:eurjmech} identified three timescales associated with parallel mean shear flow, each a measure of the time after the creation of the structure from the original disturbance:
\begin{enumerate}
\item the shear interaction timescale
\[t_s=\left[~\ov{U}_w^{'}~\right]^{-1},\]
where
\mbox{$\tau_w=\mu\ov{U}_w^{'}$}. Hence $t_s^+= 1$;
\item the viscous interaction timescale,
\[t_{\nu}=\left[L^2/(\nu\ov{U}^{'2})\right]^{1/3}, ~~t_{\nu}^+\approx20;\]
\item and the nonlinear timescale, 
\[t_n=L/u_0,~~t_n^+\approx 100;\]
\end{enumerate}
where $L$ and $u_0$ are the streamwise length scale and velocity scale, respectively, associated with the initial disturbance.  Note that here $t_n^+ \gg t_{\nu}^+$, while conventional turbulence timescales require $l/\um{} \ll l^2/\nu$ where $Re= \um{} l/\nu$ is large.  For short times after the creation of a structure from the original disturbance, the effects of both viscosity and nonlinearity may be neglected; in particular, the nonlinearity is assumed to operate only during short intermittent bursts of a local, secondary instability. Neglecting viscosity and linearising gives the Rayleigh equation for disturbances to parallel inviscid flow,
\begin{equation}
\frac{\overline{D}\nabla^2 \um{2}}{D t}-\Um^{''}\frac{\partial \um{2}}{\partial \x{1}}=0.
\label{rayleigh}
\end{equation}

\begin{figure}
\begin{center}
\includegraphics[width=8.0cm]{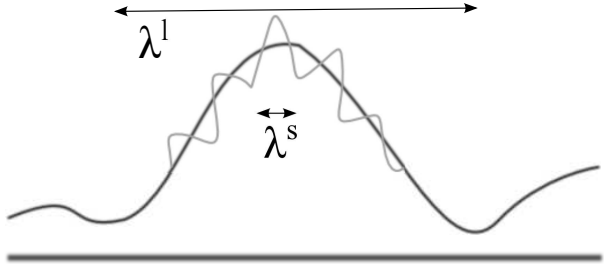}
\caption{Two-scale model of near-wall turbulence showing inner-outer interaction (after Landahl\cite{landahl75:siam}). The small-scale wavelength is denoted by $\scale{}^s$ and the larger scale by $\scale{}^l$.} \label{fig:twoscale}
\end{center}
\end{figure}

In addition to the three timescales, Landahl described two scales of motion as important in understanding turbulent shear flow. Decomposing the velocity field into large- and small-scale components,
\begin{equation}
\um{i}=\um{i}^l+\um{i}^s,
\label{decomp}
\end{equation}
the motions at the wavelengths of the large and small scales, $\scale{}^l$ and $\scale{}^s$ respectively, are assumed not to interact if $\epsilon=\scale{}^s / \scale{}^l \ll 1$.  A schematic of this description is given in Figure~\ref{fig:twoscale}.  Substitution of (\ref{decomp}) into Eqs.(\ref{orrsom}) and (\ref{squire}) and retaining only terms of leading order in $\epsilon$ provides a pair of equations, one each for the large-scale field and the small-scale field of the form of (\ref{orrsom}) and (\ref{squire}).  Later we discuss the relative importance of the different timescales and length scales to flow control and find these concepts useful even at relatively low turbulent Reynolds numbers. Landahl's scale separation has much in common with the ideas of Townsend\cite{a2t:jfm61,a2t:book76} and Bradshaw\cite{pb:jfm67} concerning inactive motion and inner-outer interaction (Morrison\cite{morrison:philtrans}).

This description also provides a basis for a more formal analysis in which wavenumber-frequency spectra are dominated by the least-damped Orr-Sommerfeld waves near resonance (see Landahl\cite{landahl67:jfm}).  For waves to be identifiable, it is necessary to form the wavenumber-frequency spectrum from which an unambiguous convection velocity can be obtained: then wave motion will appear as a reasonably narrow convective ridge\cite{wills64}.
The response may further be localised at a wall-normal location and is stronger for certain mode shapes\cite{McKeon10}.
Bark\cite{bark75} has identified the wave-like structure of the near-wall $\um{2}-$component motion as arising from the Orr-Sommerfeld eigenvalues, while some eigenvalues relating to the the horizontal components correspond to the viscous decay of wall-parallel motion.  Both Landahl\cite{landahl75:siam} and Russell and Landahl\cite{landahl84:pof} note that horizontal pressure gradients are small during these ``quiescent'' periods, and much smaller than those associated with a lift-up.

The large- and small-scale decomposition raises the question of resonance.  Jang, Benney and Gran\cite{jbg86} have proposed that, if the forcing function occurs at a frequency/wavenumber combination that matches the leading eigenmodes of the Squire equation, then resonant forcing occurs and there is the potential for large growth in amplitude before viscous damping occurs.  They have shown that such a resonance could occur at a spanwise wavenumber corresponding to a streak spacing of $90$ wall units.  Zaki and Durbin\cite{zaki05:jfm} have discussed resonance in the context of the spatial problem and have shown that the dispersion relation for the homogeneous Squire operator is identical to that of the Orr-Sommerfeld equation, making resonance possible.  However, Hultgren and Gustavsson\cite{gustavsson81:pof} have noted that since this growth mechanism is associated with the continuous spectrum, it is only possible when the flow is semi-bounded. Since our study is for a closed flow, we only encounter discrete modes.
McKeon and Sharma offer an interpretation of resonance in terms of pseudospectra\cite{McKeon10}. Essentially, they understand this resonance as the high (but non-singular) system response to harmonic forcing resulting from left half-plane (stable) eigenvalues approaching the imaginary axis. Truly neutral or inviscid modes would be located at the imaginary axis. This high response is manifested as a high resolvent norm.

\section{Model formulation}

This section describes the model formulation. We consider a three-component velocity field perturbation $\ul{}(\x{},t)$ about an assumed time-independent solution $\Ul$ in the presence of a divergence-free, bounded exogenous disturbance forcing. This gives the net velocity vector field
\begin{equation}
\label{perturb}
U_i = U_{0i} + \ul{i}.
\end{equation}
The steady pressure is similarly perturbed by $\pl(\x{},t)$.

In the next section, we will seek a control function $f(x,t)$ to globally stabilise an assumed time-independent solution $\Ul$ in the presence of a divergence-free, bounded exogenous disturbance forcing $d(x,t)$, representing unmodelled disturbances such as that arising from vibration, thermal disturbance, etc. This solution may or may not be stable to small perturbations in the absence of control. The turbulent mean is, in general, \emph{not} a time-independent solution.
Substitution of \eqref{perturb} into the Navier-Stokes equations gives the perturbation equations
\begin{subequations}
\begin{align}
\pderiv{\ul{i}}{t} =& \Pi \Big(- U_{0j}\pderiv{\ul{i}}{\x{j}}
 - \ul{j} \pderiv{U_{0i}}{\x{j}} \nonumber\\
 &+ n_i
 - \pderiv{\pl}{\x{i}}
 + \nu \pderivsq{\ul{i}}{\x{j}} + B_{ij}f_j + d_i \Big), \\
n_i =&- \ul{j} \pderiv{\ul{i}}{\x{j}},\label{NLterm}.
\end{align}
\label{perturbedNS}
\end{subequations}
A substitution has been made for the nonlinear term, giving coupled linear and nonlinear equations. The pressure term is eliminated, along with the divergence equation, by the projection $\Pi$ onto the space of divergence-free functions. We do not make the linearising assumption of small perturbations.

The forcing resulting from the control is restricted by a linear operator $B(x)$, representing physical limitations on the actuation. The range of $B$ spans the space of the divergence-free body forcing arising from all possible control actions. Thus, $Bf(x,t)$ is the forcing on the fluid arising from the control function at time $t$ and position $x$. For the purposes of understanding the current simulations, we may consider $B$ as the identity when acting on the wall-normal velocity and zero otherwise.
Let $y(x,t)$ be the measurements made at time $t$, modelled by $y_i=C_{ij}u_j$, so that $C(x)$ is a linear operator mapping the flow field to $y$.

\subsection{The discretised equations}
\label{Controller synthesis}

Discretised, the equations \eqref{perturbedNS} have the state-space form
\begin{subequations}
\begin{align}
    \dot{z} =& A z + B_{1}n + B_{1}d + B_{2}f\\
y_{1} =& C_{1} z \\
y_{2} =& C_{2} z.
\end{align}
\label{ssform3}
\end{subequations}
Matrices $C_1$ and $B_1$ are only used at the synthesis stage. They give respectively ($C_1$) the flow field at the discretisation points weighted such that $y_1(t)'y_1(t)$ approximates the perturbation energy $E(t) = \int_{x\in\Omega}u_i(x,t)u_i(x,t)~dx$, and ($B_1$) the forcing on the flow field from the nonlinearity, similarly weighted. To find $C_1$ we require the mesh weighting appropriate for the discretisation chosen. For the discretised state the perturbation energy $E(t)$ is approximated by the inner product on a positive-definite matrix $R$ so that 
$E(t) \simeq z(t)^{*}Rz(t),$ approaching equality in the continuous limit. Thus we require simply $C_{1}^{*}C_{1} = R$.
The input matrix $B_{1}$ associated with the forcing from the nonlinearity $n$ is determined similarly.
The matrix $C_2$ simply gives the measurements $y_2$ from the flow field state $z$, and is the discrete approximation of operator $C$. Similarly, $B_2$ is a matrix approximating operator $B$, describing the effect of the actuation on the flow field.

State-space representations of the linearised Navier-Stokes equations are well known in the literature (see Bewley\cite{Bewley98} for a pedagogical example). However, our formulation \emph{explicitly retains the nonlinearity as a forcing}.

\section{Controller specification and design}
This section specifies the requirements on the controller. 
For the following we will make use of the temporal-spatial inner product
\begin{equation}
\label{inprod}
\inprod{a}{b}=\int_{0}^{T}\int_{x\in \Omega} a_i^{*}(x,t) b_i(x,t) ~dx~dt
\end{equation}
and the purely spatial inner product
\begin{equation}
\label{spatialinprod}
[a,b]=\int_{x\in \Omega} a_i^{*}(x,t) b_i(x,t) ~dx.
\end{equation}

The pair $a$ and $b$ is passive if $\inprod{a}{b}\geq 0$.

We define the perturbation energy as the \emph{spatial} $L_2$ norm (induced by \eqref{spatialinprod}) of perturbations from the laminar profile, and the turbulence kinetic energy as the spatial $L_2$ norm of the perturbations from the turbulent mean profile.

The rate of change of the total perturbation energy $E$ of viscous shear flow integrated over a closed domain $\Omega \subset \bspace{R}^3$ is given by the Reynolds-Orr equation, which in parallel shear flow is
\begin{equation}
\frac{dE}{dt}=-\int_{x\in\Omega}\left(\Ul'(y) \ul{1}\ul{2} + \frac{1}{Re}D\right)~dx,
\end{equation}
where $\Ul'(y)$ is the $y$-derivative of the laminar profile $\Ul(y)$ and $D$ is the dissipation rate. As before, $x$ denotes a point in $\Omega$.
In essence, the aim of the controller is to provide actuation such that
$\frac{dE}{dt}<0$ for any disturbance.

The broad design objectives of stability and robustness are achieved by application of the passivity theorem, which gives general, open-loop conditions for closed-loop stability of two arbitrary elements connected in a feedback loop.

\subsection{The nonlinear and pressure terms}
We will use the fact that the nonlinear term \eqref{NLterm} is passive, specifically that $\inprod{u}{n} = 0$ for all $T$. Applying the divergence theorem and the divergence-free condition, it is easily shown that the inner integral in this expression is equivalent to an integral over the boundary $\partial\Omega$,
\begin{equation}
\label{nonlinbc}
\int_{x\in\Omega} \ul{i}n_{i} ~dx = \int_{x\in \partial\Omega} (\ul{j}(x,t)\ul{j}(x,t)) \ul{i}(x,t) \hat{\xi}_i ~dx
\end{equation}
where $\hat{\xi}$ is the outward-facing unit vector perpendicular to the boundary of the flow domain.

Physically interpreted, (\ref{nonlinbc}) quantifies the net flux of disturbance energy out of the domain through the boundary per unit time.
In a closed or periodic domain, the contribution to this integral from volume forcing is necessarily zero. However, in an open domain, or with transpiration at the boundary, the flux of disturbance energy through the inlet and outlet boundaries and the net rate of flux of disturbance energy from any boundary control would both contribute. Were there such a contribution, (\ref{nonlinbc}) would enter as a nonlinear constraint on the control law. For the open-domain case there will be a net flux out of the domain of the disturbance energy, where the outflow of perturbation energy is greater than in inflow of the perturbation energy (i.e.~with relatively quiet inflow conditions). In these cases, the nonlinearity has a stabilising influence in the domain of study.
In our case of periodic flow with body forcing, however, the integral is zero.

Similarly, the contribution due to the pressure term is given by,
\begin{equation}
\int_{x\in\Omega} \ul{i} \pderiv{\pl}{\x{i}} ~dx = \int_{x \in \partial \Omega} p \ul{i} \hat{\xi}_i ~dx =0.
\end{equation}
Again, in the case of actuation by boundary transpiration or in an open domain, this term would contribute to the perturbation energy. In our case, the integral is zero, simplifying the projection onto a divergence-free basis.

\subsection{The closed-loop linear terms}
\begin{figure}
    \centering
    \includegraphics{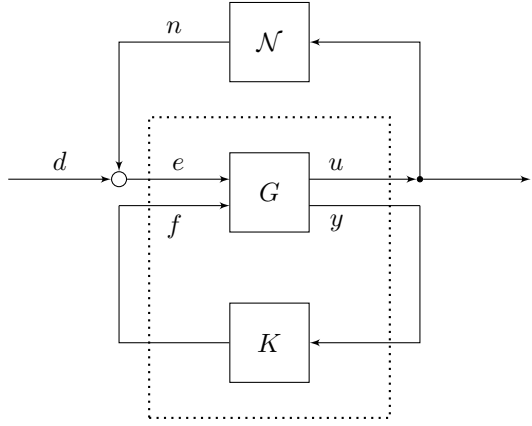}
\caption{The feedback loop for controlled Navier-Stokes. The system inside the dashed box is $Q$.}
\label{FBcontrol}
\end{figure}

In this section we will consider the discretised system equations  \eqref{ssform3} (the development is almost identical for the original perturbation equations \eqref{perturbedNS}).
We can write the discretised equations in the compact matrix notation
\begin{equation}
\left[ \begin{array}{c} \dot{z} \\ y_1 \\ y_2 \end{array} \right]
 = \left[ \begin{array}{c|cc}
A & B_{1} & B_{2} \\
\hline C_{1} & 0 & 0\\
C_{2} & 0 & 0
\end{array}\right]
\left[ \begin{array}{c} z \\ e \\ f \end{array} \right]
\label{Gss}
\end{equation}
where we have defined $e=d+n$. Let $G$ be the linear system which, when discretised, has state-space realisation \eqref{Gss} taking $e$ and $f$ to $y_1$ and $y_2$. 

Furthermore, define $K$ as the feedback law which, when discretised, generates the control action $f$ from measurements $y_2$, with state-space realisation
\begin{equation}
\left[ \begin{array}{c} \dot{z}_K \\ f \end{array} \right]
= \left[ \begin{array}{c|c}
A_K & B_{K}  \\
\hline C_{K} & 0\\
\end{array}\right]
\left[ \begin{array}{c} z_K \\ y_2 \end{array} \right].
\label{Kss}
\end{equation}

Define $Q$ as the system which maps the forcing from the nonlinearity to the flowfield, $u=Q e$. Once discretised, $Q$ is therefore the closed-loop of \eqref{Gss} and \eqref{Kss}. This arrangement is depicted as a block diagram in Figure \ref{FBcontrol}, with $Q$ being the system inside the dashed box.
Eliminating $f$ and $y_2$, the discretised $Q$ therefore has a state-space realisation
\begin{equation}
\left[ \begin{array}{c} \dot{z} \\ \dot{z}_k-\dot{z} \\ y_1 \end{array} \right]
 = \left[ \begin{array}{cc|c}
A+B_2 C_K & B_2 C_K & B_1 \\
A_K - B_2 C_K - A + B_K C_2 & A_K - B_2 C_K & - B_2\\
\hline C_{1} & 0 & 0
\end{array}\right]
\left[ \begin{array}{c} z \\ z_K - z \\ e \end{array} \right].
\label{Qss}
\end{equation}

\subsection{Stability}
The control problem therefore is to find a control forcing $f$ such that $u(x,t)\to 0$ as $t \to \infty$, given the measurements and any exogenous bounded disturbance $d$. To do this, we must consider the stability properties of the system as a whole.
Applying the passivity theorem, if $\curly{N}$ is passive and $Q$ is strictly positive real, then the closed loop in Figure~\ref{FBcontrol} (representing the controlled Navier-Stokes equations) is internally stable and is also strictly passive.  In other words, $\inprod{u}{d}> 0$. The case $\inprod{u}{d}= 0$ would imply that the forcing $d$ acts orthogonally to $u$, and $\inprod{u}{d}> 0$ implies that $d$ acts to reduce $u$ in the feedback system. This result is simply verified; from Figure \ref{FBcontrol} and by the strict positivity of $Q$ and the passivity of $\curly{N}$,
\begin{equation}
\label{inprodeqn}
\begin{split}
\inprod{u}{d} =& \inprod{u}{e-n}\\
=& \inprod{u}{e} - \inprod{u}{n}>0.
\end{split}
\end{equation}
Note that if the uncontrolled, linearised plant is already passive, no control is required, as $u$ is already bounded.
The expression $\inprod{u}{d}$ quantifies the flow perturbation energy due to the disturbance. Physically, passivity of the controlled flow implies it only dissipates perturbation energy. Since this is true for all bounded $d$, it implies that all disturbances eventually decay.

To achieve this, we wish to find a controller $K$ such that the discretised $Q$ is strictly positive real, or equivalently, $Q(s) + Q^{*}(s)>0$ or $\inprod{e}{y_{1}}_{[0,T]}>0$ for any $T$. This synthesis problem is a fairly standard exercise in robust control theory. The details of the synthesis procedure are in Appendix \ref{appendix: synthesis procedure}. Bounds on the perturbation energy production of the controlled flow are given in Appendix \ref{appendix: bounds}.

\section{Application to periodic channel flow}
To test the controller, we consider three-dimensional perturbations to plane Poiseuille flow at $Re_{\tau}=100$, with a constant mass flux. For this arrangement, laminar flow is the state with minimal sustainable drag, once control effort has been taken into consideration \cite{Fukagata05, Bewley09}. Accordingly, we aim to sustain the laminar flow state at conditions where it would otherwise not persist. The flow domain is the space between two plates parallel in the ($\x{1},\x{3}$)-plane, at $\x{2}=\pm 1$.  $\x{1}-$ and $\x{3}-$direction periodicity is assumed. We consider actuation in the whole domain provided by simple body forcing of the wall-normal velocity which is also the measurement.

The geometry allows Fourier transform of the linearised problem in the $\x{1}$ and $\x{3}$ directions which converts the spatially continuous problem into a number of decoupled continuous problems at particular Fourier wavenumber pairs. Truncation at suitably high wavenumber ensures an ($\x{1},\x{3}$)-discrete problem with sufficient resolution. Further projection onto the Chebyshev polynomials in the wall-normal direction results in a number of linear time-invariant state-space control problems. The problem thus decouples at the (linear) synthesis stage, giving a block-diagonal $A$ matrix in the state-space formulation. The decoupled wavenumbers only interact, via the nonlinearity, at the full simulation stage. The controller synthesis problem is further simplified because control is unnecessary at the highest wavenumbers where we find that viscosity dominates and that the linearised system is already close to passive.

Given that the source of the system non-normality is the interaction of the wall-normal velocity with the shear of the base profile, it is expected that actuation of wall-normal velocity only will be sufficient for the purposes of this study.

The periodic spanwise and streamwise boundary conditions are naturally enforced by the Fourier transform. Further, any forcing is divergence-free as it is expressed in a divergence-free basis.

\subsection{Implementation}
The Reynolds number based on the target parabolic profile centre-line velocity and channel half-height was $Re=1709$ (or $\Rt=100$ based on the friction velocity). The channel width was $\frac{4}{3}\pi$ and the length was $4\pi$, which was deemed sufficient to provide accurate statistics based on previous studies\cite{Kim87}. The simulation was performed using a modified version of Channelflow~0.9.15 \cite{Gibson} which solves for the primitive variables using spectral discretisation in the spatial directions (Chebyshev in the wall-normal direction, Fourier otherwise). The flow field was advanced in time by a mixed third-order Runge-Kutta scheme, that treats the linear terms implicitly and the nonlinear terms explicitly.
The number of modes used was $32 \times 71 \times 32$.  A variable time step was set capped at $\Delta t=0.01$ which was sufficiently short to keep the  CFL number low. The nonlinear terms were computed in skew-symmetric format with 3/2 dealiasing in the wall-parallel directions.
The statistics of the unmanipulated flow were verified by comparing them to a database provided by Kuroda and Kasagi \cite{Kuroda98}. The profiles of the mean velocity and the Reynolds stresses largely collapse, with a slight discrepancy in the $\um{1}$-component normal stress near the centre-line. Given the accuracy, the discretisation is considered sufficient for this study.

The control action was integrated using a zero-order hold, stepped at $\Delta t=0.01$. The control penalty was set at $\varepsilon=0.01$ as defined in (\ref{augmentedG}). The control action was restricted to forcing on the wall-normal velocity and the sensing was likewise restricted. A value of $\gamma<1.02$ was achieved for the Cayley-transformed system at all wavenumbers, indicating that the controlled flow is very close to passive.

\begin{figure}
\begin{center}
\includegraphics[width=8.0cm]{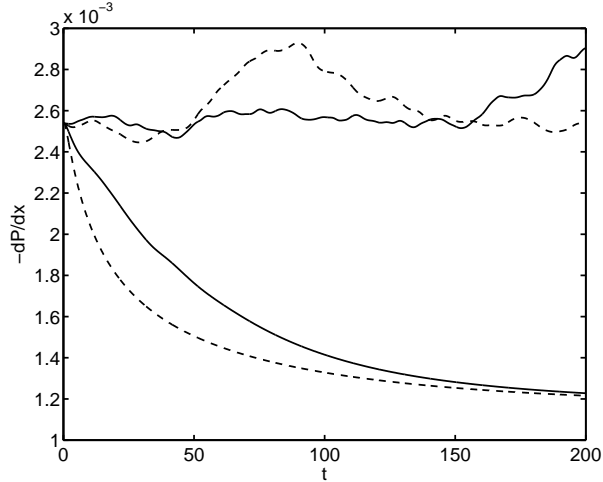}
\caption{Pressure gradient variation with time: upper (--), uncontrolled; upper (- -) $k_1, k_3\le2$; lower (- -), $k_1, k_3\le4$; lower (--), $k_1, k_3\le8$.}
\label{fig:ddpdxdt}
\end{center}
\end{figure}
Figure~\ref{fig:ddpdxdt} shows the pressure gradient for the controlled cases decreasing with time. Interestingly, the restriction of control to lower wavenumbers ($k_1, k_3\le4$) produces forcing that is almost as effective as that for the larger wavenumber range ($k_1, k_3\le8$). The control fails when restricted to $k_1, k_3 \leq 2$. This suggests that the effect of viscous damping is significant enough at the highest wavenumbers to overcome the energy production due to the shear interaction. Consequently, we infer that the dominant production mechanisms occur at these larger scales: the key requirement is that the control scheme should resolve streaks and streamwise vortices.  Hence, additional control at $k_1, k_3 > 8$ is ineffective because the forcing appears at scales shorter than the streak spacing ($k_1 \simeq 8$ corresponds to $\scale{3}^+ \simeq 75$).
\begin{figure}
\centering
{\includegraphics[width=8.0cm]{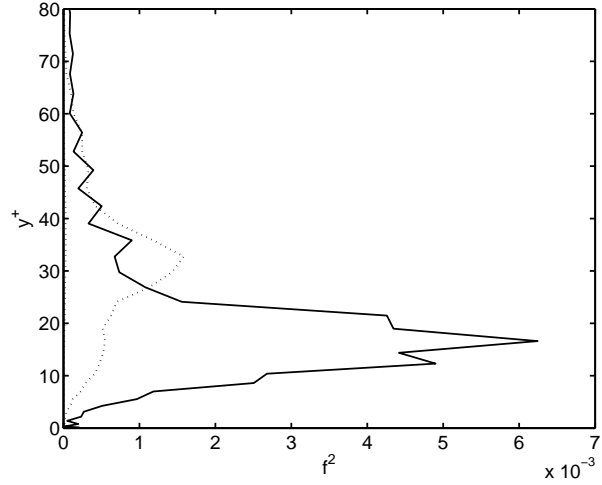}}
\caption{Mean-square forcing (averaged over wall-parallel planes) at $t=10$ (--) and $t=50$ ($\cdot \cdot \cdot$). The forcing decreases substantially over time. It peaks close to $\x{2}^+\approx 20$.}
\label{fig:forcing_profile}
\end{figure}

\begin{figure}
\centering
\subfigure[~forcing at  $\x{2}\simeq 0$]
{\includegraphics[width=8.0cm]{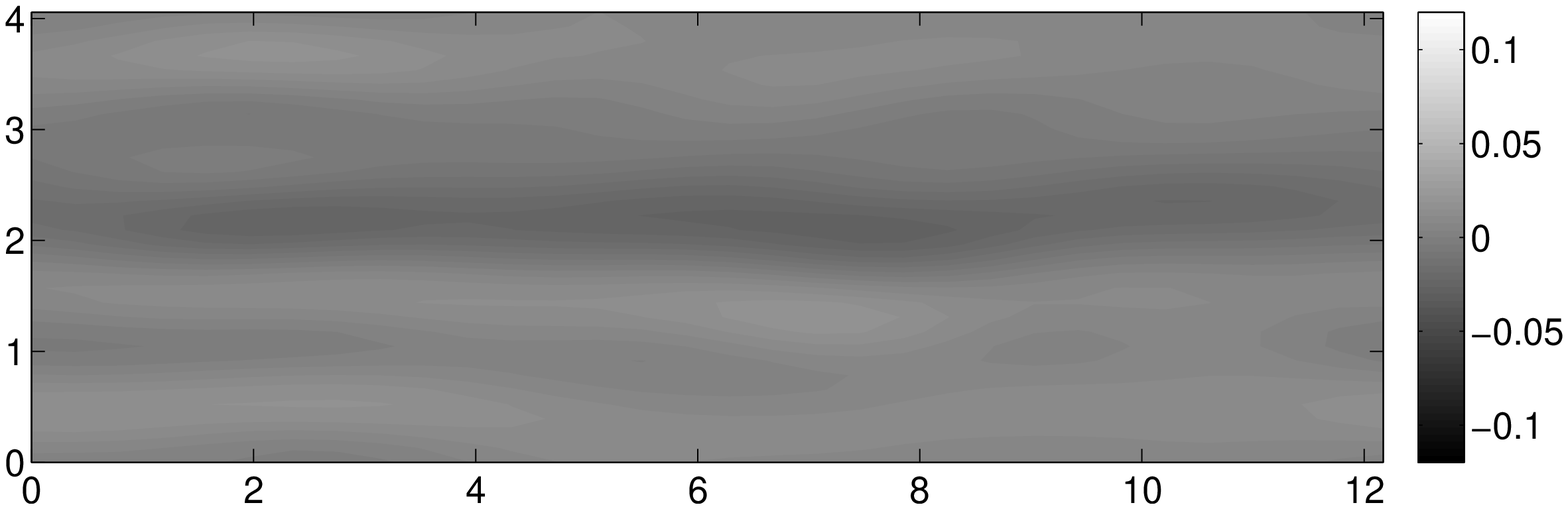}}\\
\subfigure[~forcing at $\x{2}^+=20$]
{\includegraphics[width=8.0cm]{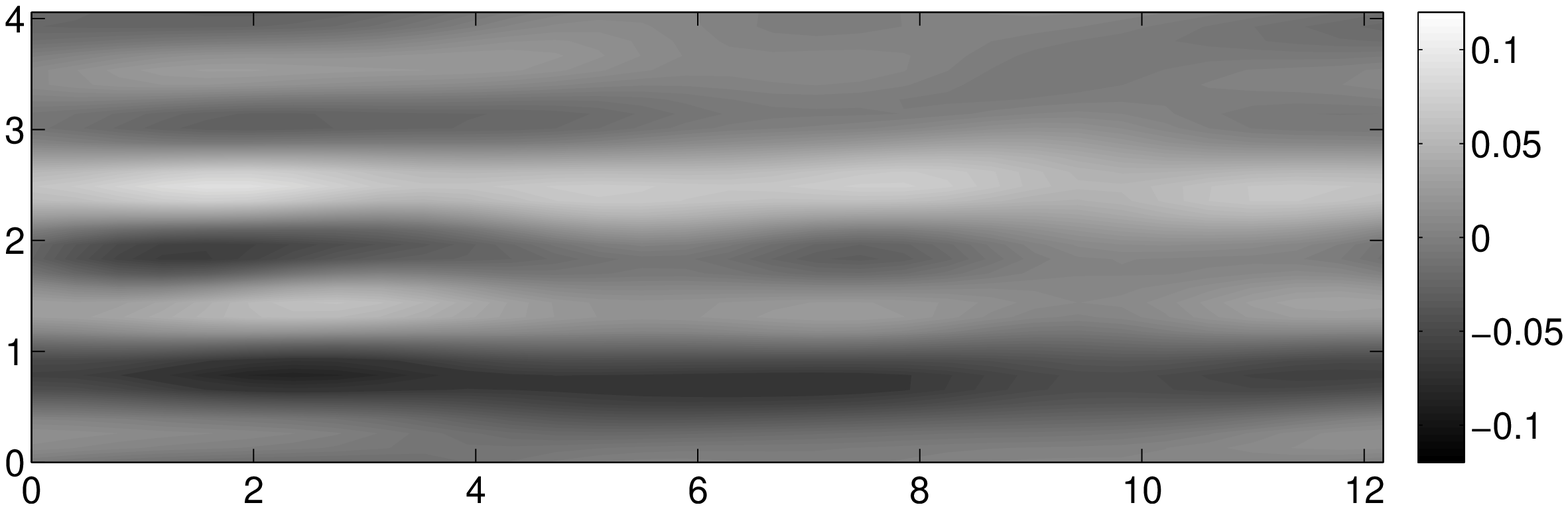}}\\
\subfigure[~forcing at $\x{2}=h$]
{\includegraphics[width=8.0cm]{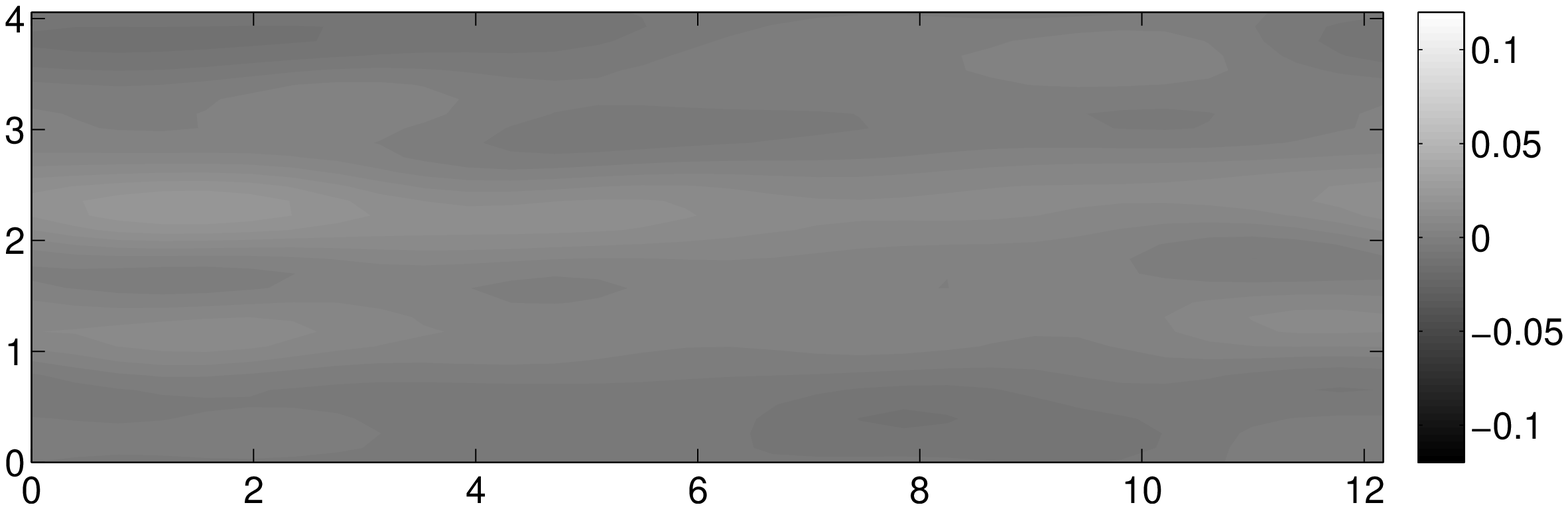}}\\
\caption{Contours at various wall-normal distances of the forcing provided by the controller at $t=10$}
\label{fig:forcing}
\end{figure}

\begin{figure}
\centering
\includegraphics[width=8.0cm]{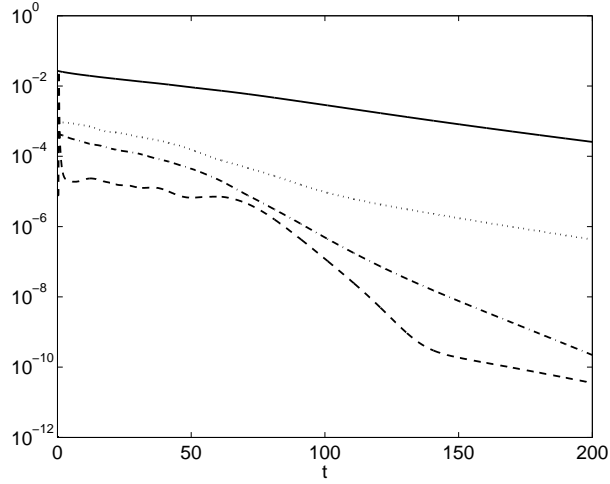}
\caption{Evolution of the average of $u_1^2$ (--), $u_2^2$ ($-\cdot$), $u_3^2$ ($\cdot\cdot\cdot$) and $p^2$ (-~-) with control at $k_1, k_3\le4$. The log scale shows that the rate of decline is fastest for $\ul{2}$ and $p$.}
\label{fig:norms_evl}
\end{figure}

\begin{figure}
\centering
\subfigure[~contours of $\ul{1}$ at $\x{2}^+=20,~t=10$]
{\includegraphics[width=8.0cm]{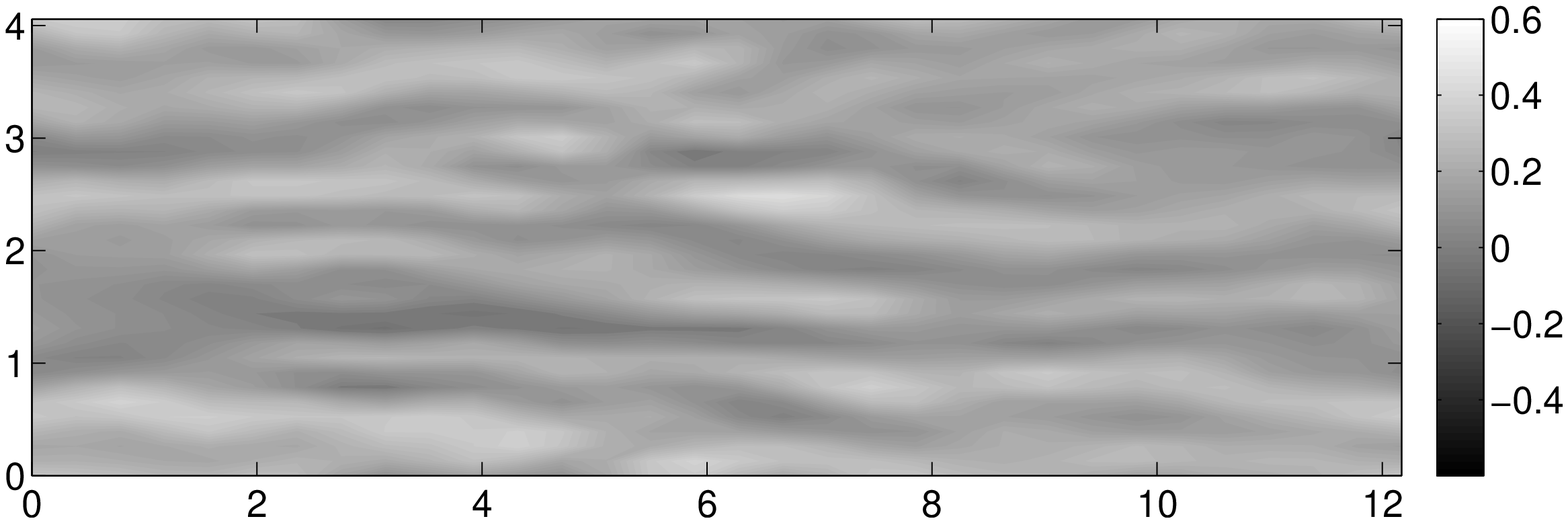}}\\
\subfigure[~contours of $\ul{1}$ at $\x{2}^+=20,~t=20$]
{\includegraphics[width=8.0cm]{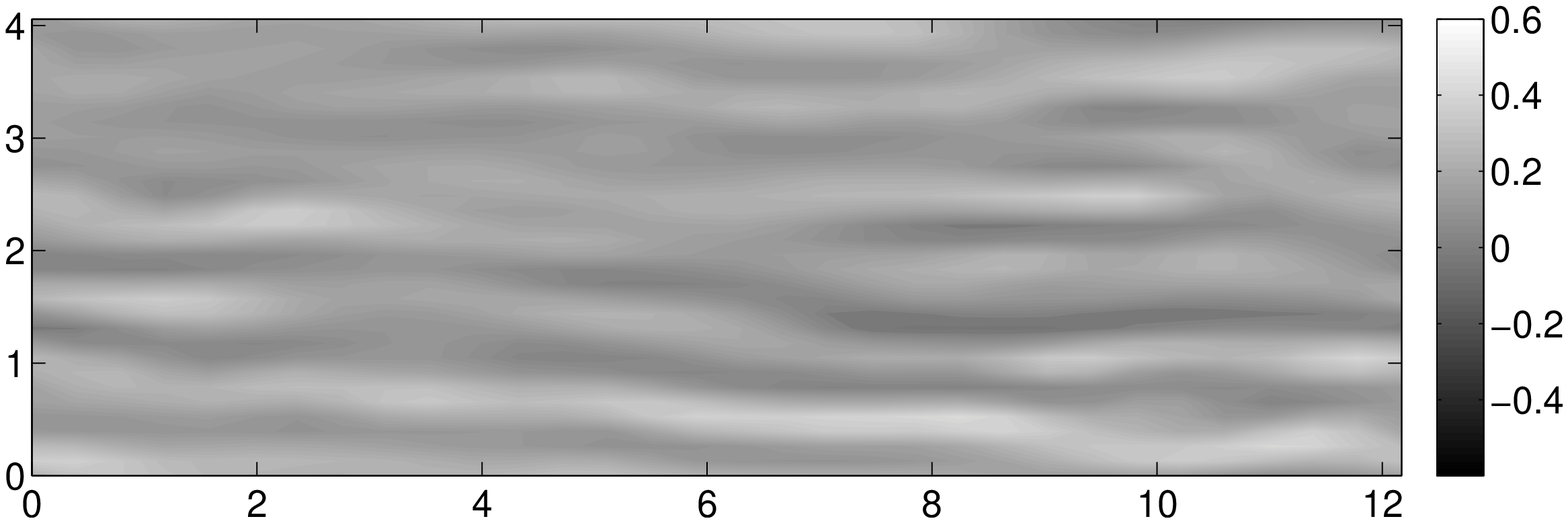}}\\
\subfigure[~contours of $\ul{1}$ at $\x{2}^+=20,~t=50$]
{\includegraphics[width=8.0cm]{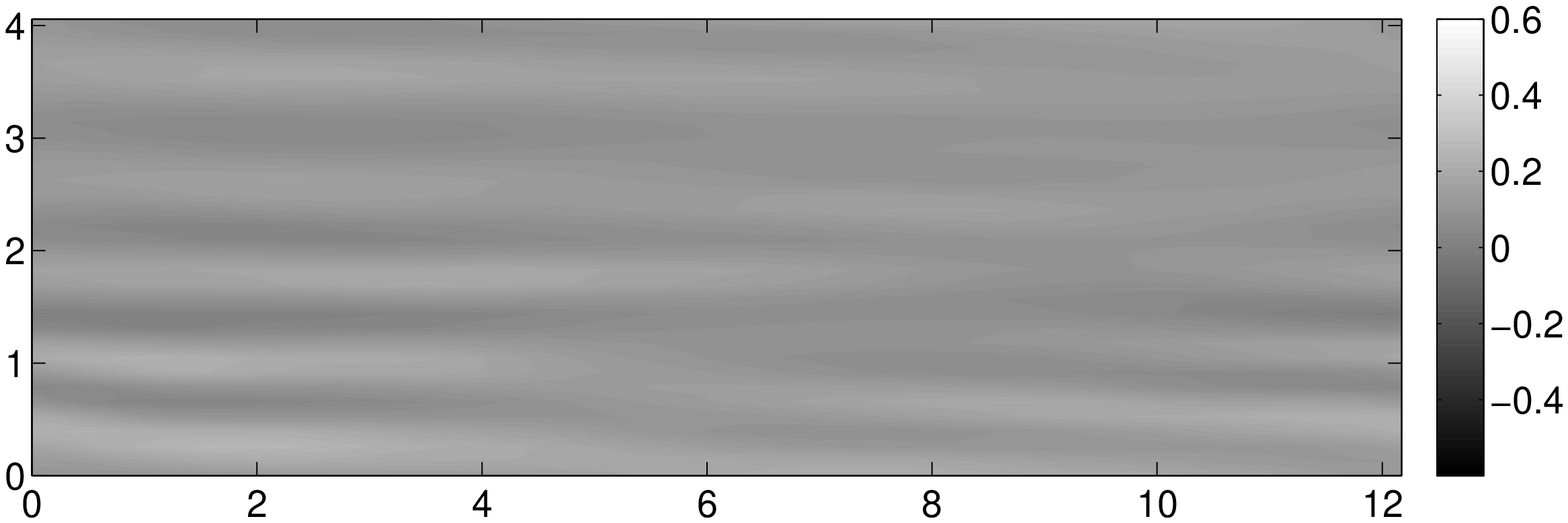}}
\caption{Contours of $\ul{1}$ for the controlled flow at $\x{2}^+=20,~t=10,~20,~50$}
\label{fig:contours_u}
\end{figure}

\begin{figure}
\centering
\subfigure[~contours of $\ul{2}$ at $\x{2}^+=20,~t=10$]
{\includegraphics[width=8.0cm]{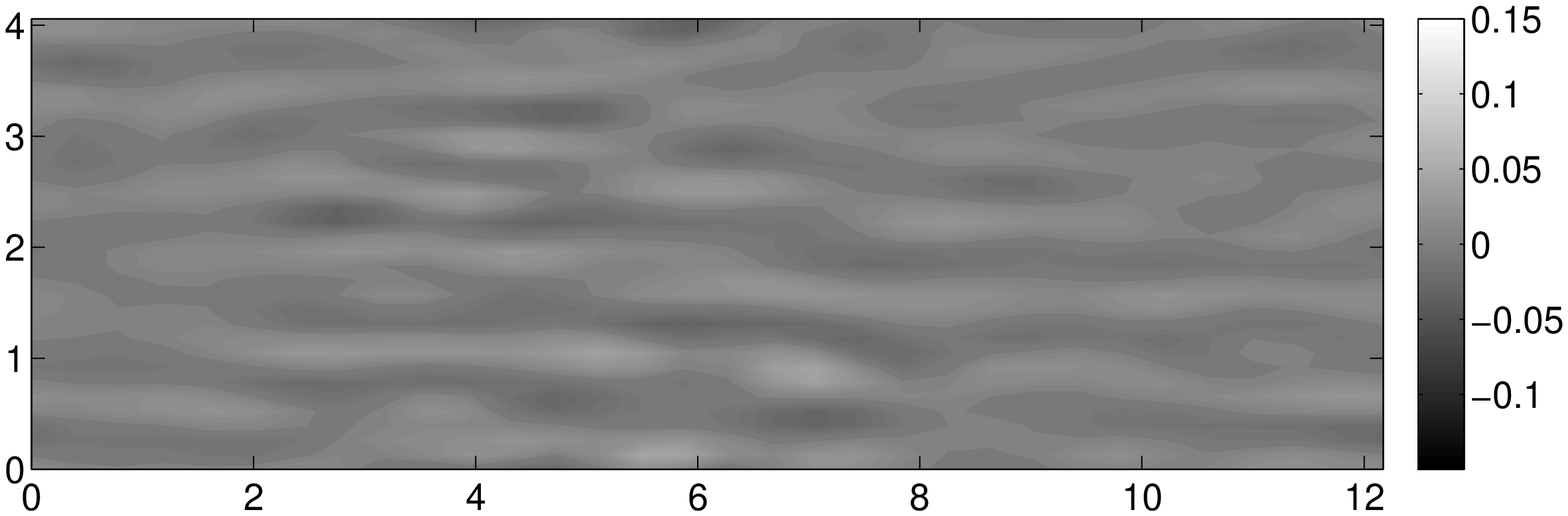}}\\
\subfigure[~contours of $\ul{2}$ at $\x{2}^+=20,~t=20$]
{\includegraphics[width=8.0cm]{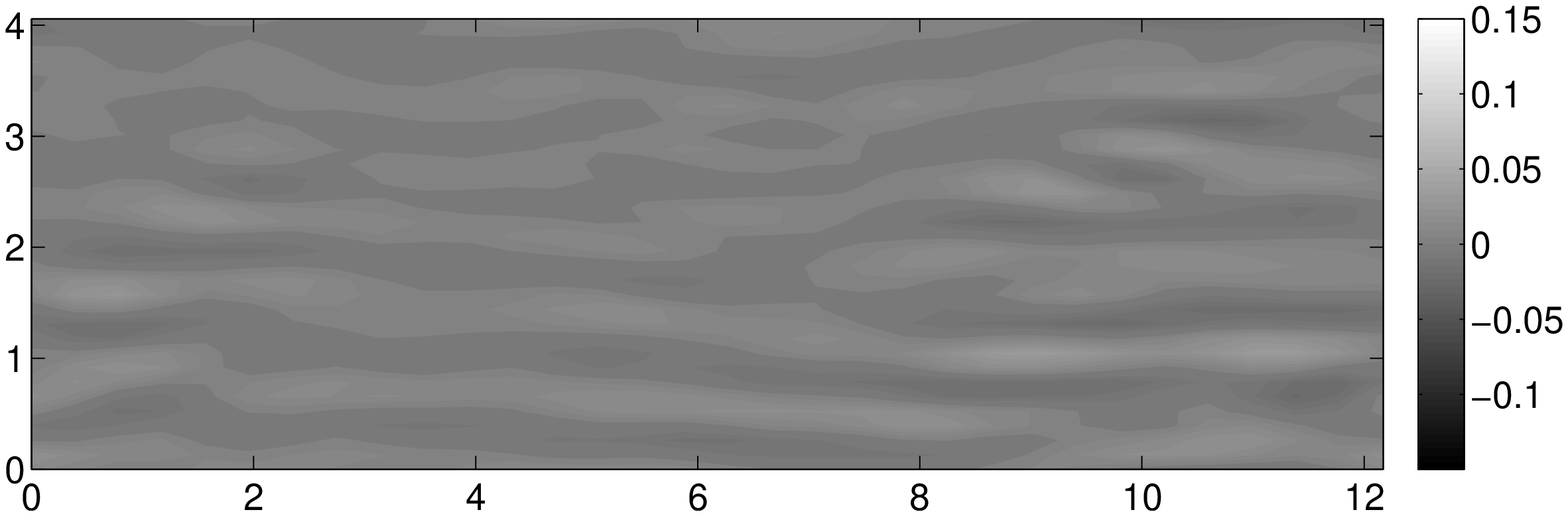}}\\
\subfigure[~contours of $\ul{2}$ at $\x{2}^+=20,~t=50$]
{\includegraphics[width=8.0cm]{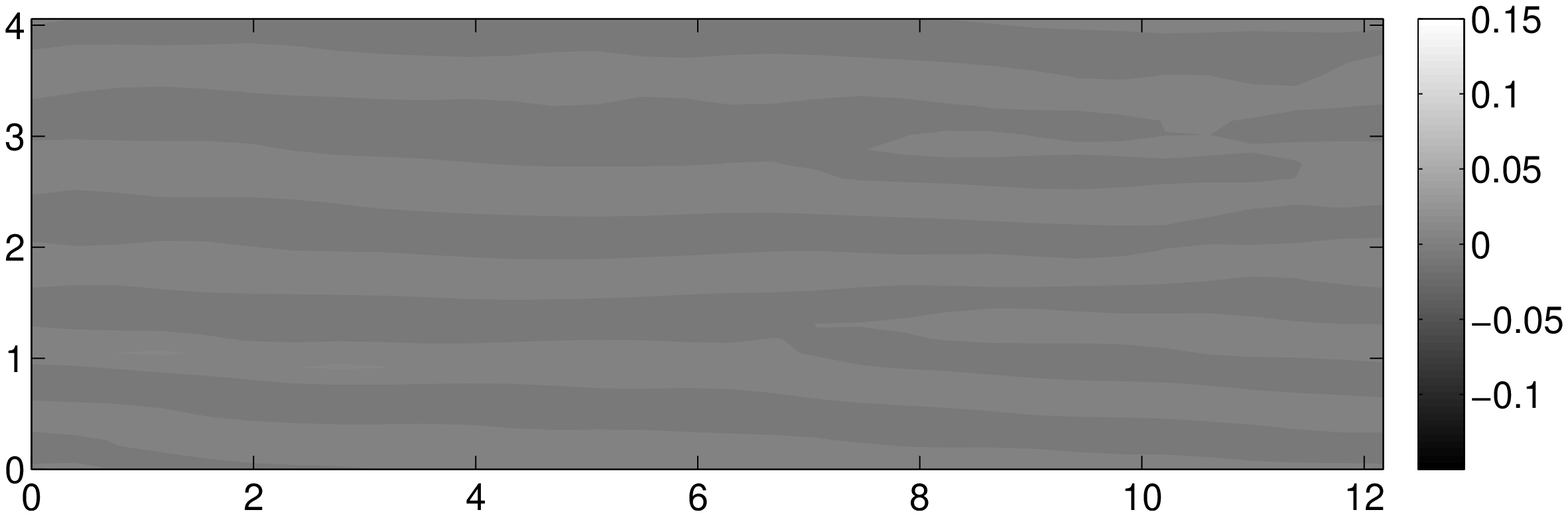}}
\caption{Contours of $\ul{2}$ for the controlled flow at $\x{2}^+=20,~t=10,~20,~50$}
\label{fig:contours_v}
\end{figure}

\begin{figure}
\centering
\subfigure[~contours of $\ul{3}$ at $\x{2}^+=20,~t=10$]
{\includegraphics[width=8.0cm]{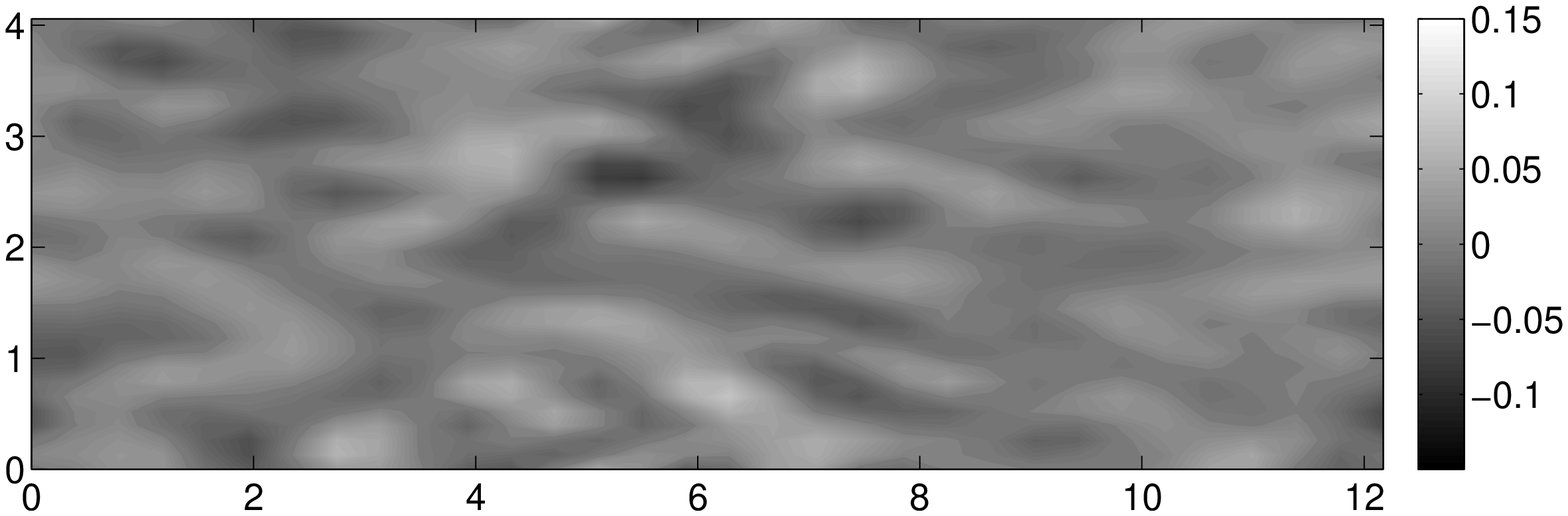}}\\
\subfigure[~contours of $\ul{3}$ at $\x{2}^+=20,~t=20$]
{\includegraphics[width=8.0cm]{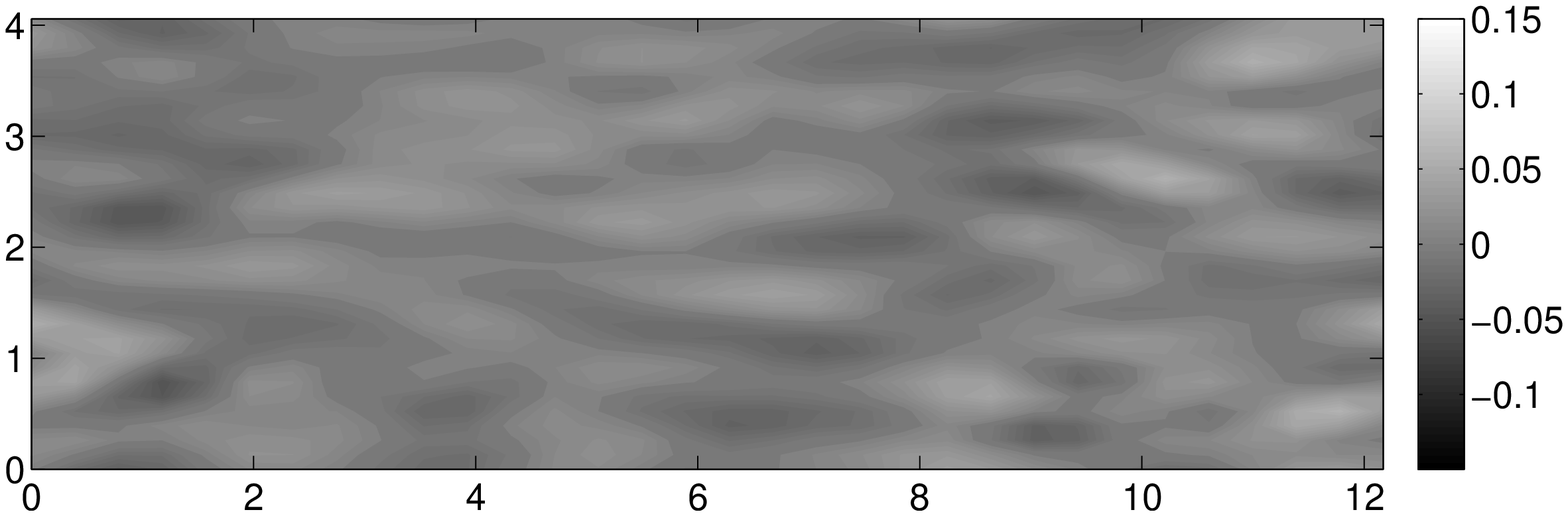}}\\
\subfigure[~contours of $\ul{3}$ at $\x{2}^+=20,~t=50$]
{\includegraphics[width=8.0cm]{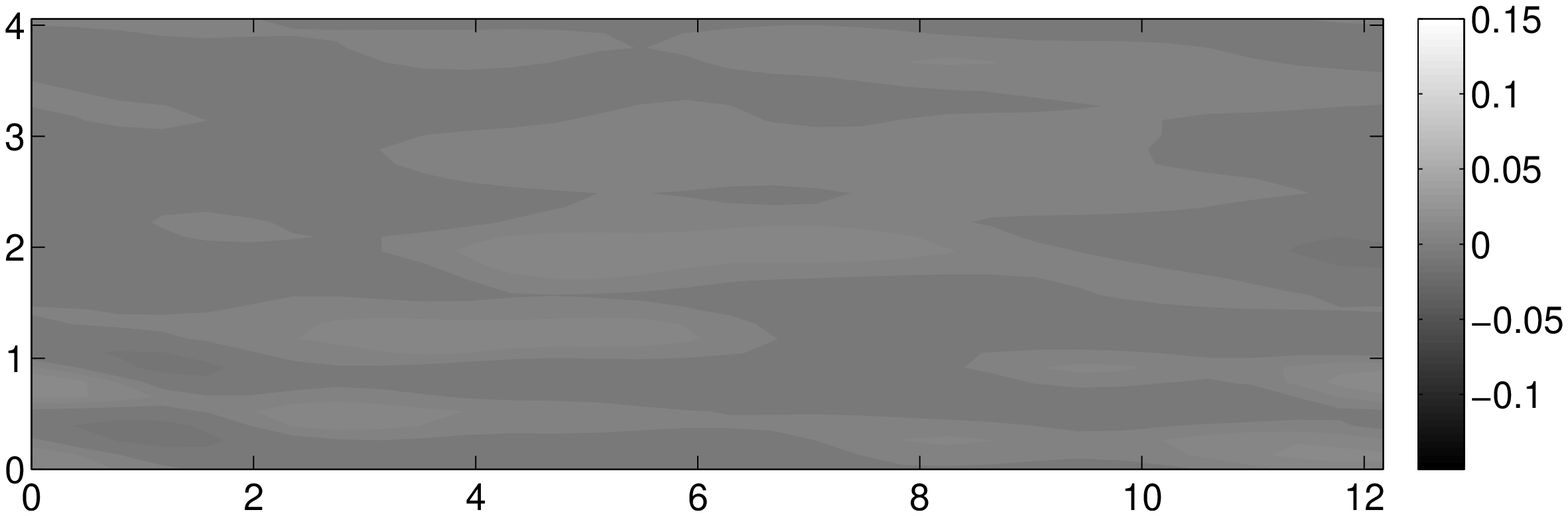}}
\caption{Contours of $\ul{3}$ for the controlled flow at $\x{2}^+=20,~t=10,~20,~50$}
\label{fig:contours_w}
\end{figure}

\begin{figure}
\centering
\subfigure[~contours of $p$ at $\x{2}^+=20,~t=10$]
{\includegraphics[width=8.0cm]{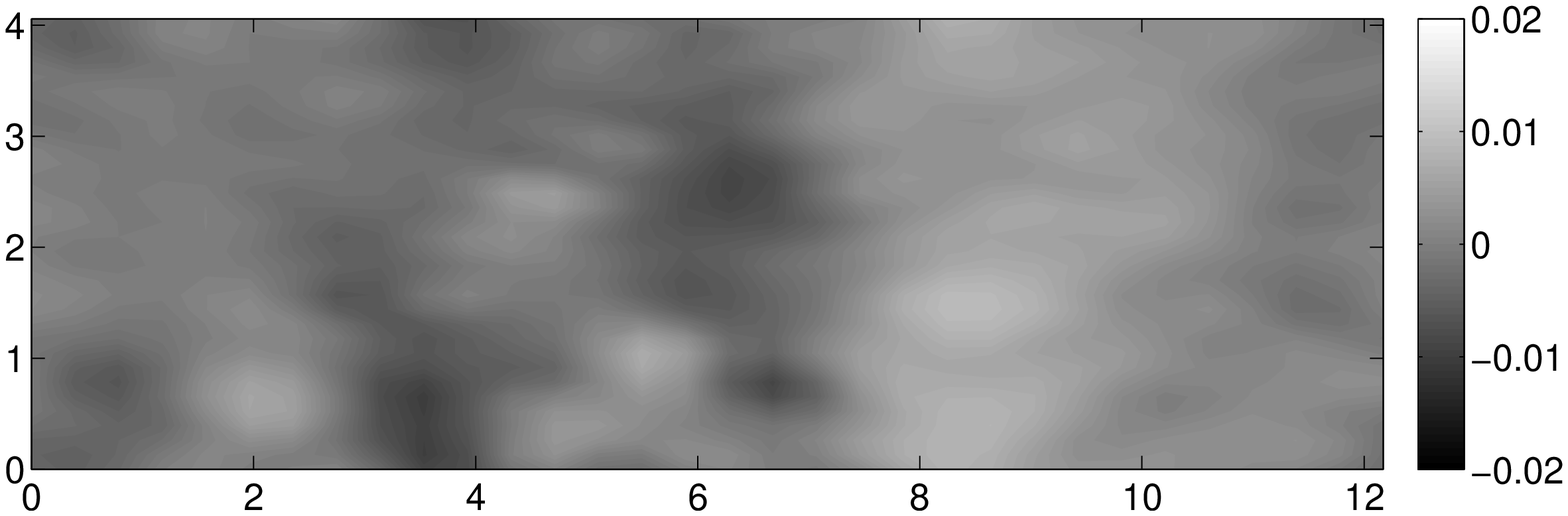}}\\
\subfigure[~contours of $p$ at $\x{2}^+=20,~t=20$]
{\includegraphics[width=8.0cm]{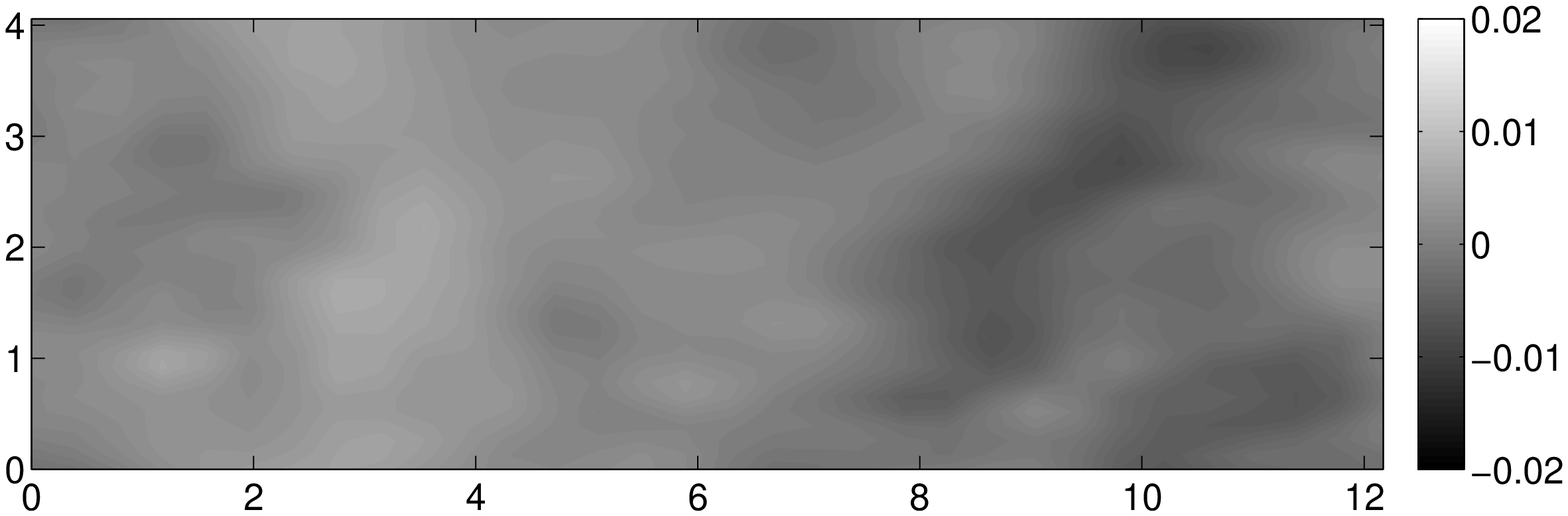}}\\
\subfigure[~contours of $p$ at $\x{2}^+=20,~t=50$]
{\includegraphics[width=8.0cm]{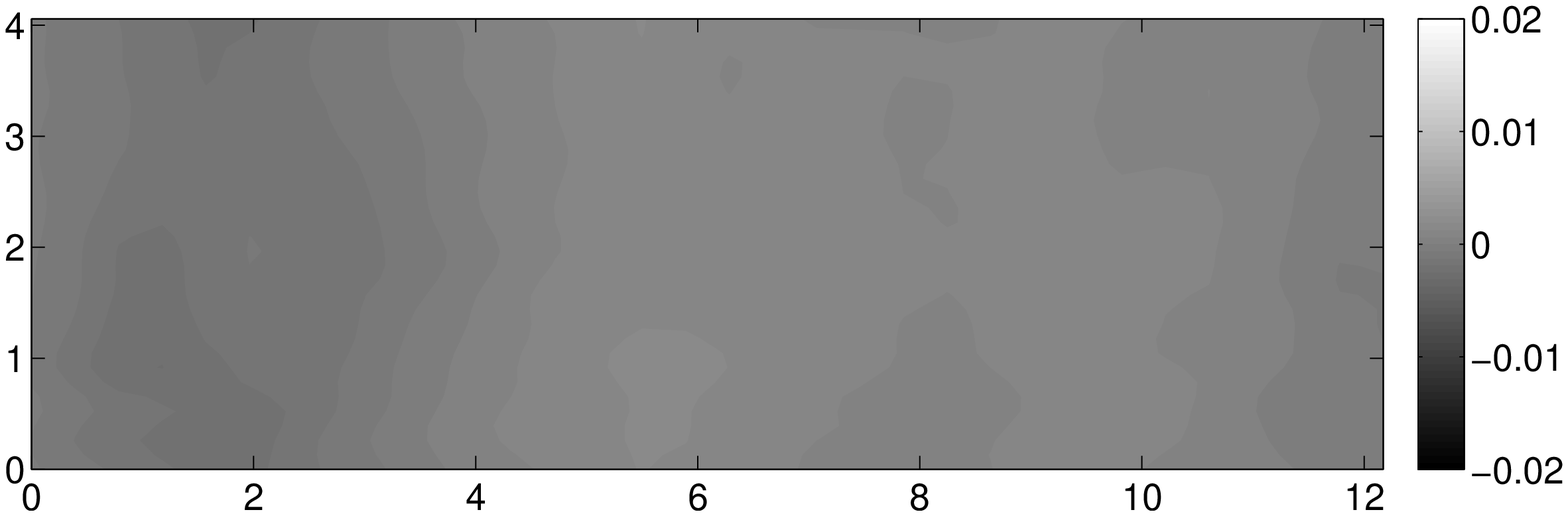}}
\caption{Contours of $p$ for the controlled flow at $\x{2}^+=20,~t=10,~20,~50$}
\label{fig:contours_p}
\end{figure}

Figure~\ref{fig:forcing_profile} shows the mean-square averages over wall-parallel planes of the forcing as it varies with wall-normal distance at various times. Where variables are expressed in viscous units ({\it e.g.} $\x{2}^+$), the relevant viscous scale is calculated from the uncontrolled flow (otherwise it would change with time). The forcing is concentrated around $\x{2}^+\simeq 20$ where the shear interaction is most significant. The forcing peak decreases over time and moves further into the flow interior. This indicates that, as the laminar profile is approached, only minimal control effort is required. Contours of the forcing at various wall-normal distances are shown in Figure~\ref{fig:forcing}.

Next we examine the results in relation to the pressure field.
Figure \ref{fig:norms_evl} shows the energy of the velocity and pressure fields over time, for one controlled case.
Figures \ref{fig:contours_u}, \ref{fig:contours_v}, \ref{fig:contours_w} and \ref{fig:contours_p} show contours of these fields at $\x{2}^+=20$ for various times. Comparison shows that the wall-normal and pressure perturbations are controlled very quickly. The spanwise perturbations subsequently decay, then lastly the energetic streaky streamwise contours. There is a brief spike in the pressure field as the controller comes on-line. This ordering supports the picture that the interaction of wall-normal motion and shearing is a minimum requirement for the control of streaks\cite{Chernyshenko05}.

\section{The role of pressure and wall-normal velocity fluctuations}
The Poisson equation for pressure fluctuations in reduced form appropriate for channel flow is given by
\begin{equation}
-{\nabla^2 \pmt}=2\Um'\frac{\partial \um{2}}{\partial \x{1}}-\frac{\partial ^{2}}{\partial \x{1}\partial \x{2}}\Big[ \um{1}\um{2}-\overline{\um{1}\um{2}}\Big]
\label{poisson}
\end{equation}
where we are once again considering perturbations to the turbulent mean profile.
The first term on the right-hand side is the linear or ``rapid" source and the second term is the nonlinear or ``slow" term, the physical distinction coming from the fact that the linear term changes as soon as the mean rate of strain changes.  Pressure fluctuations are known to be well correlated across a shear flow: Kim\cite{kim89} has shown that the two-point correlation extends over both large wall-normal distances and over large spanwise distances near the centre-line of turbulent channel flow.  He has also shown that contributions come mainly from the slow source term, except close to the wall where contributions from the rapid and slow terms are about the same.  The earlier measurements of Sternberg\cite{sternberg62} also indicate that the linear pressure fluctuation field at the edge of the sublayer is larger than the nonlinear field.

The rapid term arises from the inviscid, linear term, associated with the interaction of the wall-normal disturbances with the mean profile. The proposition that this is the leading term that must be controlled, is consistent with the controller's success, and the localisation of the control near $\x{2}^+\simeq 20$. This accepted, the control may reasonably be expected to work when restricted to actuation and sensing on wall-normal velocity or pressure disturbances.
Landahl\cite{landahl65:nasa} has formulated the perturbation field in terms of the pressure as:
\begin{equation}
\frac{\overline{D}\nabla^2 \pmt}{D t}-2\Um^{'}\frac{\partial^2 \pmt}{\partial \x{1}\partial \x{2}}-\frac{\nabla^4 \pmt}{Re}=s,
\label{porrsom}
\end{equation}
where $s$ is the nonlinear term corresponding to that in equation (\ref{orrsom}).  Interpretation of the the terms on the left-hand side suggests that terms like $\nabla^2 \pmt$ and $\nabla^4 \pmt$ will be significant where $\pmt$ changes rapidly.

\begin{figure}
\centering
\subfigure[~reproduced from Dunn and Morrison\cite{dunn:jfm}]
{\includegraphics[width=8.0cm]{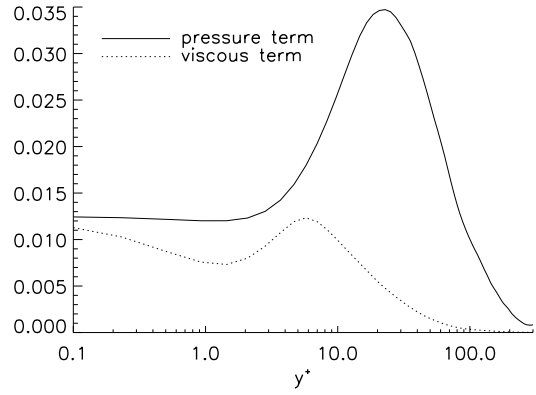}}\\
\subfigure[~for $t=0$]
{\includegraphics[width=8.0cm]{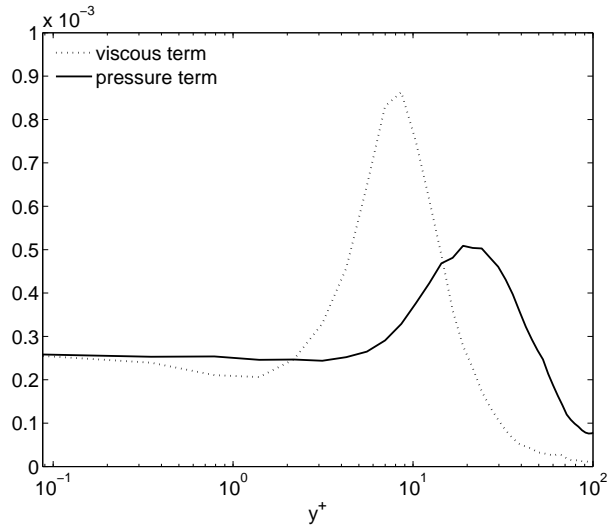}}\\
\subfigure[~for $t=50$, controlled flow]
{\includegraphics[width=8.0cm]{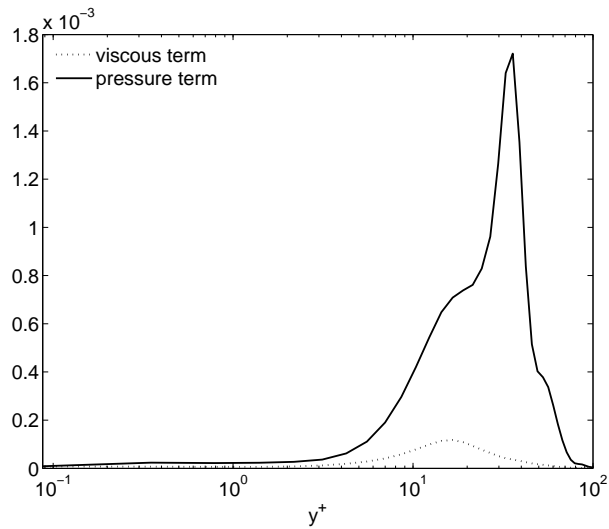}}
\caption{Mean-square pressure gradient and mean-square viscous force, defined by Equation~(\ref{batch2}), but with the lower two figures using the target laminar profile, not the mean profile. The peak of the pressure term occurs at $\x{2}^+\simeq20$ until the control acts.}
\label{fig:force terms}
\end{figure}

\begin{figure}
\centering
{\includegraphics[width=8.0cm]{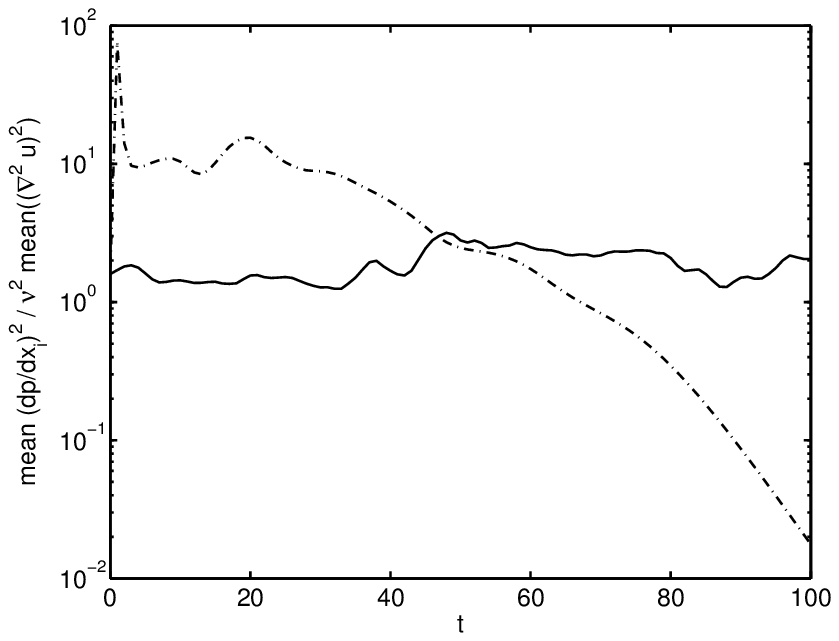}}
\caption{Evolution of the ratio of the mean-square pressure perturbation gradient to the mean-square perturbed viscous terms at $\x{2}^+=20$ (uncontrolled (--), controlled ($-\cdot$) at $k_x, k_z\le4$).}
\label{fig:meansq_evl}
\end{figure}

More physically, Kim\cite{kim89} shows that, in channel flow at $Re_{\tau}=179$, the static pressure is only slightly negatively skewed, but has flatness factors that are typically twice the Gaussian value of three over much of the channel height.  Writing the mean-square acceleration as
\begin{equation}
 \overline{\Biggl(\frac{D \um{i}}{Dt}^2\Biggr)}= \overline{\left (\frac{\partial \pmt}{\partial \x{i}}\right)^2}
 +\nu^2 \overline{\left (\frac{\partial^2 \um{i}}{\partial x^2_j}\right )^2}-2\nu\frac{\pa}{\pa x_i}\left(\overline{\pmt\frac{\pa^2 \um{i}}{\pa x^2_i}}    \right),
  \label{batch1}
\end{equation}
Batchelor and Townsend\cite{batchelor:townsend} have shown that, at Reynolds numbers high enough for local isotropy (such that the diffusion term is negligible), the mean-square pressure gradient is much larger than the mean-square viscous force.  Further, they suggest that 
\begin{equation}
\overline{\left (\frac{\partial p}{\partial x_i}\right
  )^2} \approx 20\nu^2 \overline{\left (\frac{\partial^2 u_i}{\partial
  x^2_j}\right )^2}, 
  \label{batch2}
\end{equation}
where the constant is determined empirically.
Dunn and Morrison\cite{dunn:jfm} (see Figure~\ref{fig:force terms}) show that, outside the viscous sublayer, the factor is about $5-10$, even at low Reynolds numbers.
In the current work ($Re_{\tau}=100$), we observe it to be about $2$ at $\x{2}^+ =20$ for the uncontrolled case, as shown in Figure~\ref{fig:meansq_evl}.

\begin{figure}
\begin{center}
\includegraphics[width=8.5cm]{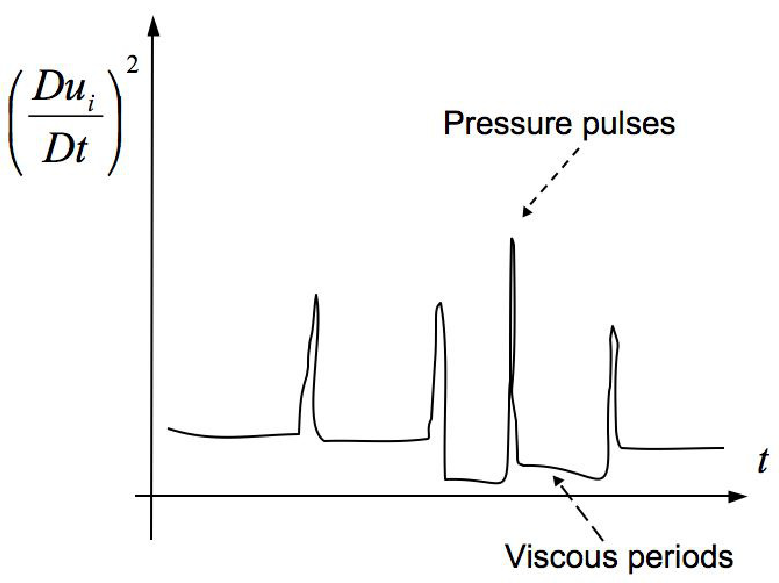}
\caption{Schematic of variation of mean-square acceleration with time: prolonged viscous periods pulsed by pressure `spikes' generated principally by quasi-streamwise vortices.}
\label{fig:pulses}
\end{center}
\end{figure}

Equation (\ref{batch1}) suggests that the mean-square acceleration comprises prolonged viscous intervals `pulsed' periodically by the mean-square pressure gradient, as illustrated in Figure~\ref{fig:pulses}.  Thus a pressure field distribution of small skewness, but large flatness, gives rise to a pressure-gradient distribution of which the first moment is very small, but with even moments that are significantly larger.  Kim\cite{kim89} also shows that contributions to the mean-square wall pressure are principally local in nature even though the instantaneous wall pressure receives significant contributions from the opposite wall of the channel.  Therefore, the mean-square pressure close to the surface is intimately related to the structure there. In terms of the sublayer populated with quasi-streamwise vortices, this means merely that a low-pressure region (approximately coinciding with the vortex core) always has two opposite-signed pressure gradients in the cross-sectional plane of the vortex. Hence, at $\x{2}^+ \approx 25$,
$\overline{(\partial \pmt/\partial \x{1})^2}\approx 0.5\overline{(\partial \pmt/
\partial \x{2})^2}\approx 0.5\overline{(\partial \pmt/\partial \x{3})^2}$.
Kim \cite{kim89} also notes that $\partial p/\partial x$ is not a good indicator of quasi-streamwise vortices whereas the vertical and spanwise gradients are.

The relevance of Landahl's equations lies in the fact that, over the short time for which the controller is active, the considerably longer turbulence timescale means that the turbulence itself is not very significant. They therefore offer an explanation of the controller's success.  The controller reduces the pressure gradients over time through action on the $v-$component and the linear source term in equation~(\ref{poisson}). In the short term ($t<35$), the pressure term is higher than in the uncontrolled case, reflecting the controller's action. In the very short term, there is a brief spike, perhaps due to the initialisation of the controller. The viscous term decreases almost monotonically (not shown). The net result is that the ratio of the pressure terms to the viscous terms increases for a short time, then declines as the controller action takes effect.  Figure \ref{fig:force terms} shows that the location of the maximum mean-square pressure gradient initially occurs at $\x{2}^+\approx 20$, where the forcing is a maximum. The controller moves the peak location of this term over time, presumably as the effective Reynolds number drops.

\section{Conclusions}
A new characterisation of stabilising feedback laws for incompressible Navier-Stokes flows has been presented in terms of passivity theory. The control is designed to make the magnitude of any perturbation to the laminar operating point decay monotonically.
The flow equations are discretised and the ensuing controller synthesis problem results in two game-theoretic algebraic Riccati equations. When these Riccati equations have solutions, a globally stabilising, linear, controller can be synthesised. A simple synthesis procedure has been presented which is derived by the Cayley transformation of the positivity problem into an auxiliary $\gamma$-optimisation \Hinf problem. Tools for the $\gamma$-optimisation problem are widely available in packages such as MATLAB\textsuperscript{\textregistered}\cite{MATLAB} or Octave\cite{Octave}.
A control effort penalty and measurement noise model has been introduced to avoid a singular control problem and its associated large control signals.
The methodology allows an attempt at control with limited or insufficient actuation or sensing and permits bounds on the maximum perturbation energy production.
It has been applied to turbulent channel flow with wall-normal interior body sensing and forcing. It was verified that the control relaminarised the flow, even when the forcing was confined to low wavenumbers ($k_x,k_z\leq4$). It seems likely that an important requirement is for the mean streak spacing to be resolved. Intuitively, this requirement explains the targeting of the shear interaction mechanism.

We have seen that controlling the wall-normal perturbations successfully, and with it the pressure perturbations, resulted in the eventual collapse of the streamwise streaky structures. This causality shows that the interaction of wall-normal motion and shearing is necessary for the formation of these streaks.

The success of the control may be understood in terms of Batchelor and Townsend's result showing the importance of pressure-gradient fluctuations, and several essential features of Landahl's model. 
We observe that the shear interaction timescale is shorter than the viscous and nonlinear (turbulent) timescales. Since the shear interaction process is essentially linear and underpins the turbulent fluctuations, our control strategy is also linear.
The shear interaction is governed by the wall-normal disturbance, which is related to the pressure via the linear (``fast") source term in the Poisson equation for pressure fluctuations. Consequently, the control may be satisfactorily restricted to wall-normal velocity or pressure.

The response of the pressure and wall-normal velocity disturbances is particularly high for large, wave-like motions at close to the convective velocity, which correlate over significant distances in planes parallel to the wall. These waves are remarkably non-dispersive, with an approximately constant phase velocity.
At higher Reynolds numbers, we might expect the effect on the near-wall turbulence of superstructures to become more important, with correspondingly more stringent requirements on actuation authority. These are also wave-like disturbances, but associated with a near-singular response of the linear terms to the nonlinear convection term\cite{McKeon10} which may be understood in relation to critical layer theory.

In Landahl's theory, when the phase velocity is equal to the group velocity of disturbances emanating from further upstream, a secondary instability in the form of a burst occurs\cite{landahl72:jfm,landahl77:pof}. Inhibiting the propagation of these waves precludes the occurrence of such nonlinear secondary instabilities associated with turbulent flow. Again, this aspect to the control problem is essentially linear and inviscid.

Thus we may interpret the control via Landahl's theory; the linear mechanisms on the ``rapid'' timescale are most important, which is why the controller concentrates on the shear instability. The controller concentrates on regions where this mechanism is most active, acting to ``normalise'' the response of the system. This can be achieved through manipulation of just the wall-normal component of the velocity field.

\begin{acknowledgments}
We are grateful to EPSRC for financial support under grant number EP/E017304. AS also wishes to acknowledge the support of an Imperial College Junior Research Fellowship. BJM wishes to acknowledge NSF award number 0747672 (program manager H. Henning Winter). We are grateful to Adelaide de Vecchi for assistance with the channel simulations, to Ahmed Naguib and Tamer Zaki for insightful and happy discussions.
\end{acknowledgments}

\appendix


\section{The passivity control synthesis procedure}
\label{appendix: synthesis procedure}

\subsection{Overview}

\begin{figure*}[t]
\begin{center}
\includegraphics[width=\textwidth]{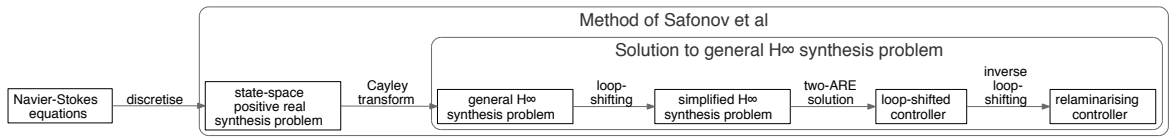}
\caption{A flow diagram giving an overview of the entire controller synthesis procedure.}
\label{fig:flow-diagram}
\end{center}
\end{figure*}

This appendix details the process of finding a discrete controller that satisfies the closed-loop passivity requirement for the discretised system.

The aim is simply to provide the reader with enough information to replicate the procedure. No proofs are presented, and lengthy explanation is avoided. Such proofs are available in the book by Green \& Limebeer\cite{Green}, which our presentation follows closely. That text, alongside Kailath\cite{Kailath}, serves as a good background reference on the systems theory and robust control theory used in this work.

This appendix is split up into a number of sections. The first section gives an overview of the synthesis procedure. The second section presents the transformation of the positive real synthesis problem into a general \Hinf problem. Then, the third section gives the loop-shifting transformations needed to convert the general problem into a simplified problem. Finally, in the fourth section, the solution to the simplified \Hinf control problem is given.

\subsection{Overview}
There are many possible approaches to solving the passivity control problem. 
In the approach chosen for this work, the synthesis problem is solved by applying a transformation to the system we wish to make passive, giving a new control problem where we have to find a controller to make the infinity norm of closed-loop with the transformed plant less than $1$ \cite{Safonov87}.
The resulting general \Hinf problem is in turn solved using loop-shifting transformations \cite{Green, Safonov88} and Riccati-based state space methods \cite{Green, Doyle89}. The multi-step process is outlined in Figure \ref{fig:flow-diagram}.

The chosen approach is not necessarily the simplest (for instance, that of Sun \cite{Sun94} is more direct), however it is robust and enables utilisation of readily available software such as the MATLAB\textsuperscript{\textregistered} robust control toolbox\cite{MATLAB}.

Four assumptions are made. Of these assumptions, one requires the stabilisability and detectability of the uncontrolled system. This is automatically satisfied if the flow is below the critical Reynolds number, where the first unstable eigenvalue appears.
A second assumption is imposed to prevent unbounded control signals. The remaining assumptions are required for the solution method of the Riccati equations, and may be relaxed.

The iterative method presented is useful in the case where there is insufficient actuation or sensing to make the closed-loop passive. In this case, we iteratively search for a controller to get the system as close to passive as is possible.
This relaxation loses the strict guarantee of nonlinear stability, but it is still possible to quantifiably limit the perturbation energy production. We choose this method for our study, because it is more amenable to such a relaxation and so may be more applicable to cases with physical or design constraints on the available measurement and actuation.

The solution to the general \Hinf problem is somewhat intricate, however it can be simplified using loop-shifting transformations summarised in Section \ref{appendix:loop shifting}, so that the simplified \Hinf theory presented in Section \ref{appendix:simplified synthesis} can be applied. This synthesis method requires the solution of two Algebraic Riccati Equations (AREs) at each wavenumber.

For brevity, the notation in each appendix is self-contained.

\subsection{Transformation of the positive real synthesis problem to a general \Hinf synthesis problem}
\label{appendix:passivity synthesis}

Let $G$ have state-space matrices given by
\begin{equation}
G = \left[ \begin{array}{c|cc}
A & B_{1} & B_{2} \\
\hline C_{1} & 0 & 0\\
C_{2} & 0 & 0
\end{array}\right].
\label{G}
\end{equation}

The closed-loop transfer function of $G$ and a controller $K$ will be strictly positive real if and only if the closed-loop transfer function of $\Gt$ and $K$ has infinity norm less than $1$ (see Safonov {\it et al.}\cite{Safonov87}), where 
\begin{equation}
\tilde{G} = \left[ \begin{array}{c|cc}
A - B_{1}C_{1}  & {B}_{1} & {B}_{2} \\
\hline -2{C}_{1} & I & 0\\
{C}_{2} & 0 & 0
\end{array}\right].
\label{Gtilde}
\end{equation}

The problem therefore has become to find a controller $K$ to minimise the \Hinf norm, $\gamma$, of the closed-loop of $K$ with $\Gt$. If $\gamma<1$, the closed-loop between the original system $G$ and $K$ is strictly positive real, thereby solving the original passivity problem. There is no \emph{a priori} way to find a minimal $\gamma$, so it is necessary to perform an iterative search over $\gamma$.

\subsubsection{Control penalty}

The control problem as presented above permits unbounded control signals, essentially because it does not penalise the control effort. This is tackled by introducing a penalty on the control, and a model for sensor noise.
The penalties are made orthogonal to the dynamics, by augmenting $\tilde{G}$ (to give $\tilde{G}^{+}$), with scalar the penalty weighting $\varepsilon$,
\begin{equation}
\tilde{G}^{+} = \left[ \begin{array}{c|cc}
A - B_{1}C_{1}  &
{\left[ \begin{array}{cc} B_{1}&  0 \end{array} \right]} & {B}_{2} \\
\hline
{\left[ \begin{array}{c} -2{C}_{1}\\ 0 \end{array} \right]}  & {\left[ \begin{array}{cc} I & 0\\0&0 \end{array} \right]}  & {\left[ \begin{array}{c} 0 \\ \varepsilon I \end{array} \right]}\\
{C}_{2} & {\left[ \begin{array}{cc} 0 & \varepsilon I \end{array} \right]}  & 0
\end{array}\right].
\label{augmentedG}
\end{equation}
It will be seen that the penalty is necessary to satisfy rank assumptions on $D_{12}$ and  $D_{21}$ of Section \ref{appendix:loop shifting}.

At this point, we apply the results of sections \ref{appendix:loop shifting} and \ref{appendix:simplified synthesis} to find the required controller.

We find that the control penalty is the primary obstacle to minimising $\gamma$.

\subsection{The loop shifting transformations}
\label{appendix:loop shifting}
This section describes, without explanation, the loop shifting transformations required to convert the problem into a form that is solved for in Section \ref{appendix:simplified synthesis}.

From the previous section, we have a system of the form
\begin{equation}
P(s) =
\mat{c|cc}{
	A & B_1 & B_2 \\
	\hline
	C_1 & D_{11} & D_{12} \\
	C_2 & D_{21} & D_{22}
}
\end{equation}
with $A\in\mathbb{C}^{n\times n}$, $B_1\in\mathbb{C}^{n\times m}$, $C_2\in\mathbb{C}^{q\times n}$, and other matrices dimensioned accordingly.

The \Hinf controller synthesis formulae are greatly simplified by assuming that $D_{11}=0$ and $D_{22}=0$ and that $D_{12}$ and $D_{21}$ satisfy some simple rank assumptions.

Let the state-space matrices satisfy the following assumptions,
\begin{enumerate}[{A}1.]
\item $(A,B_2,C_2)$ is stabilisable and detectable,
\item $\rank (D_{12})=m$ and $\rank (D_{21})=q$
\item $\rank \mat{cc}{
	j\omega I - A & -B_2 \\
	C_1 & D_{12}
}
=m+n$ for all real $\omega$
\item $\rank \mat{cc}{
	j\omega I - A & -B_1 \\
	C_2 & D_{21}
}=q+n$ for all real $\omega$
\end{enumerate}

The aim is to replace the system $P$ with an equivalent problem involving $\hat{P}$, where
\begin{equation}
\Ph(s) =
\mat{c|cc}{
	\Ah & \Bh_1 & \Bh_2 \\
	\hline
	\Ch_1 & 0 & \Dh_{12} \\
	\Ch_2 & \Dh_{21} & 0
}
\end{equation}
with the simplified assumptions,
\begin{enumerate}[$\mathrm{\hat A}$1.]
\item $(\Ah,\Bh_2,\Ch_2)$ is stabilisable and detectable,
\item $\Dh_{12}^* \Dh_{12} = I_m$ and $\Dh_{21} \Dh_{21}^* = I_q$
\item $\rank \mat{cc}{
	j\omega I - \Ah & -\Bh_2 \\
	\Ch_1 & \Dh_{12}
}
=m+n$ for all real $\omega$
\item $\rank \mat{cc}{
	j\omega I - \Ah & -\Bh_1 \\
	\Ch_2 & \Dh_{21}
}=q+n$ for all real $\omega$
\end{enumerate}

\subsubsection{Minimise $||\hat{D}_{11}||$}
We define
\begin{widetext}
\begin{align}
\Pb(s) &= \mat{c|cc}{
	A + B_2 F (I-D_{22}F)^{-1}C_2  &  B_1 + B_2 F (I-D_{22} F)^{-1} D_{21}  & B_2 (I-F D_{22})^{-1} \\ \hline
	C_1 + D_{12} F (I-D_{22}F)^{-1} C_2  &  D_{11} + D_{12} F(I-D_{22} F)^{-1} D_{21}  &  D_{12} (I-F D_{22})^{-1}\\
	(I-D_{22}F)^{-1}C_2  &  (I-D_{22}F)^{-1} D_{21}  &  (I-D_{22} F ) ^{-1} D_{22}
} \nonumber \\ &= 
\mat{c|cc}{
	\Ab & \Bb_1 & \Bb_2 \\ \hline
	\Cb_1 & \Db_{11} & \Db_{12} \\
	\Cb_2 & \Db_{21} & \Db_{22}
}
\end{align}
\end{widetext}

Begin by choosing $F$ such that $||\Db_{11}||=\gamma_0$ is minimised, where $\gamma_0 = \max\{||\Dh_{12}^* D_{11}||,||D_{11}\Dh_{21}^*||\}$. This can be done in more that one way (see Green \& Limebeer\cite{Green} for further details).

\subsubsection{Eliminate $\Dh_{11}$}
Define 
\begin{equation}
\mat{cc}{
\Theta_{11} & \Theta_{12} \\
\Theta_{21} & \Theta_{22}
} = \gamma^{-1}
\mat{cc}{
	\gamma^{-1}\Db_{11} & (I-\gamma^{-2}\Db_{11}\Db_{11}^*)^{1/2} \\
	-(I-\gamma^{-2} \Db_{11}^* \Db_{11})^{1/2} & \gamma^{-1} \Db_{11}^*
}.
\end{equation}
We can eliminate $\Dh_{11}$, by substitution we see directly that

\begin{widetext}
\begin{align}
\Ph(s) &= \mat{c|cc}{
	\Ah + \Bh_1 \Theta_{22}(I-\Dh_{11}\Theta_{22})^{-1}\Ch_1  & \Bb_1(I-\Theta_{22}\Db_{11})^{-1}\Theta_{21}  & \Bb_2+\Bb_1\Theta_{22}(I-\Db_{11}\Theta_{22})^{-1}\Db_{12}\\\hline
	\Theta_{12}(I-\Db_{11}\Theta_{11}\Theta_{22})^{-1}\Cb_1 & 0 & \Theta_{12} (I-\Db_{11} \Theta_{22})^{-1} \Db_{12} \\
	\Cb_{2} + \Db_{21} \Theta_{22} (I- \Db_{11} \Theta_{22})^{-1} \Cb_1  & \Db_{21}(I-\Theta_{22}\Db_{11})^{-1} & \Db_{22} + \Db_{21}\Theta_{22}(I-\Db_{11}\Theta_{22})^{-1}\Db_{12}
} \nonumber \\ &=
\mat{c|cc}{
	\Ah & \Bh_1 & \Bt_2\\ \hline
	\Ch_1 & 0 & \Dt_{12} \\
	\Ct_2 & \Dt_{21} & \Dh_{22}
}.
\end{align}
\end{widetext}

\subsubsection{Eliminate $\Dh_{22}$}
Eliminate $\Dh_{22}$ by connecting $-\Dh_{22}$ in parallel with $\Ph_{22}$.

\subsubsection{Rank conditions on $\Dh_{12}$ and $\Dh_{21}$}

Find scaling matrices $S_1$ and $S_2$ such that $\Dh_{12}=\Dt_{12}S_1$ with $\Dh_{12}^* \Dh_{12}=I_m$, and similarly $\Dh_{21}=S_2\Dt_{21}$ with $\Dh_{21} \Dh_{21}^*=I_q$. The rescaled system is then
\begin{equation}
\Ph(s) = \mat{c|cc}{
	\Ah & \Bh_1 & \Bh_2\\\hline
	\Ch_1 & 0 & \Dh_{12} \\
	\Ch_2 & \Dh_{21} & 0
}
\label{Phat}
\end{equation}

\subsubsection{Controller synthesis}
\label{appendix:synthesis}
Find the controller $\Kt$ to solve the small gain problem for the system $\Ph$ in \eqref{Phat}, using the method presented in Appendix \ref{appendix:simplified synthesis}.

\subsubsection{Reversing the loop shifting}
The final step is to apply the preceeding steps of this appendix to the controller $\Kt$ 
 in reverse.
Where 
\begin{equation}
\Kt=\mat{c|c}{
	\At_k & \Bt_k \\ \hline
	\Ct_k & 0
},
\end{equation}
the final controller is given by

\begin{widetext}
\begin{equation}
K=\mat{c|c}{
	\At_k + \Bt_k S_1 \Dh_{22} (I+F \Dh_{22})^{-1}S_2 \Ct_k & \Bt_k S_1 - \Bt_k S_1 \Dh_{22} (I-D\Dt_k \Dh_{22})^{-1} F \\ \hline
	S_2 \Ct_k -F \Dh_{22} (I+F\Dh_{22})^{-1}S_2 \Ct_k & -F + F\Dh_{22}(I+F\Dh_{22})^{-1}F
}.
\end{equation}
\end{widetext}

\subsection{Solution to the simplified \Hinf control synthesis problem}
\label{appendix:simplified synthesis}

In this section we present the controller synthesis formulae solving the small gain problem of Section \ref{appendix:synthesis} above.

Suppose the system $P$, given by
\begin{equation}
P(s) = \mat{c|cc}{
	A & B_1 & B_2\\\hline
	C_1 & 0 & D_{12} \\
	C_2 & D_{21} & 0
},
\end{equation}
satisfies the simplified assumptions of Section \ref{appendix:loop shifting}. We seek a controller $K$ such that the closed-loop of $P$ and $K$ is stable and the infinity norm of the closed-loop is less than $\gamma$.

There exists such a $K$ if and only if
\begin{enumerate}
\item There exists a solution $X$ to the ARE \eqref{RicX} such that $\At - (B_2 B_2'-\gamma^{-2}B_1 B_1')X$ is asymptotically stable and $X\geq 0$.
\item There exists a solution $Y$ to the ARE \eqref{RicY} such that $\Ab - Y(C_2 C_2'-\gamma^{-2}C_1 C_1')$ is asymptotically stable and $Y\geq 0$.
\item The spectral radius, $\rho(XY)<\gamma^2$.
\end{enumerate}

The AREs in question are
\begin{equation}
\label{RicX}
X\At + \At'X - X(B_2 B_2' - \gamma^{-2}B_1 B_1')X + C_1'(I-D_{12}D_{12}')C_1 = 0
\end{equation}
with $\At=A-B_2 D_{12}' C_1$, and

\begin{equation}
\label{RicY}
\At Y + Y\At' - Y(C_2' C_2 - \gamma^{-2}C_1' C_1)Y + B_1(I-D_{21}'D_{21})B_1' = 0
\end{equation}
with $\Ab=A-B_1 D_{21}' C_2$.

When these conditions are met, one such controller is given by 
\begin{equation}
K=\mat{c|c}{
	A_k & B_k\\ \hline
	C_k & 0 }
\end{equation}
with
\begin{align*}
A_k &=A + \gamma^{-2}B_1 B_1'X - B_2 D_{12}'C_1+B_2'X \\
    &\qquad+ B_k (C_2 + \gamma^{-1}D_{21}B_1'X),\\
B_k &= B_1 D_{21}' Y(I-\gamma^{-2}XY)^{-1} (C_2 + \gamma^{-1}D_{21}B_1'X)',\\
C_k &=-D_{12}'C_1-B_2'X.
\end{align*}

\section{Bounds on the perturbation energy}
\label{appendix: bounds}
The solution of the auxiliary small-gain problem ($\gamma < 1$) results in monotonic decay of the disturbance energy. In the case that $\gamma \geq 1$, this property may be lost, however the method does optimise for the \emph{worst-case} perturbation energy production.
This is seen from the following argument.

A transfer function $Q$ is strictly positive real, if and only if its Cayley transform $\Qt$ has infinity norm less than $1$, \ie $\norm{Q}{\infty}<1$ \cite{Safonov87}.
The Cayley transform $\tilde{Q}$ of system $Q$ is given by
\begin{equation}
\tilde{Q}(s) = (Q(s) - I)(Q(s) + I)^{-1}.
\label{Cayley}
\end{equation}

We have transformed the problem of making some transfer function $Q(s)$ as close as possible to positive real into an equivalent problem of making $\tilde{Q}(s)$ bounded real, \ie $\norm{\tilde{Q}}{\infty} < \gamma$.
Then
\begin{equation}
\det [I - \gamma^{-1} \tilde{Q}(s)] \neq 0, \quad \mathrm{for~} \real s >0.
\end{equation}
Using the Cayley transform (\ref{Cayley}) it is straightforward to show that
\begin{align}
\tilde{Q}(s)&\tilde{Q}^{*}(s)\nonumber\\
=& (Q(s)-I)(Q(s)+I)^{-1} (Q^{*}(s)+I)^{-1}(Q^{*}(s)-I) \nonumber\\
\leq & \gamma^{2}I.
\end{align}
Rearrangement gives
\begin{equation}
Q(s) + Q^{*}(s) \geq \frac{1-\gamma^{2}}{1+\gamma^{2}} (Q^{*}(s)Q(s) + I).
\end{equation}
As $\gamma \rightarrow 1$, $Q(s)$ becomes positive real.
Bounding the right hand side by $-\alpha$,
\[-\alpha = \inf_{s=j\omega} \left[ \frac{1-\gamma^{2}}{1+\gamma^{2}} (Q^{*}(s)Q(s) + I) \right] \]
means $\alpha \geq 0$ (since $\gamma \geq 1$).

If $u=Qe$, then it is straightforward to show
\begin{equation}
\inprod{u}{e} \geq - \frac{\alpha}{2} \inprod{e}{e} ~ \forall e.
\end{equation}
Since $\norm{u}{2}$ is the perturbation energy, this bounds the rate of perturbation energy production by any disturbance $e$ and optimising $\gamma$ optimises this bound.

\end{document}